\pdfoutput=1
\documentclass[a4paper,11pt]{article}

\usepackage{lineno}
\usepackage{jheppub}
\usepackage[utf8]{inputenc}
\usepackage{amsmath}
\usepackage{amsthm}
\usepackage{amssymb}
\usepackage{enumitem}   
\usepackage[english]{babel}
\usepackage{url}
\usepackage{mathtools}
\usepackage{bbold}
\usepackage{slashed}
\usepackage{multirow}
\usepackage{lipsum}
\usepackage{xcolor}
\usepackage{float}
\usepackage{revsymb}
\usepackage{braket}
\usepackage{siunitx}

\allowdisplaybreaks

\newcommand{\bm}[1]{\boldsymbol{#1}}

\newcommand{\curly}[2]{\mathcal{{#1}}^{({#2})}}
\newcommand{\barcurly}[2]{\bar{\mathcal{{#1}}}^{({#2})}}
\newcommand{\tildecurly}[2]{\tilde{\mathcal{{#1}}}^{({#2})}}
\newcommand{\D}{\mathrm{d}}
\newcommand{\p}{\partial}

\newcommand{\rot}{\mathrm{rot}}

\newcommand{\diver}{\mathrm{div}}

\definecolor{dgreen}{rgb}{0,0.6,0.0}

\newcommand{\gc}{e}

\begin{document}


\preprint{MS-TP-24-20}
\title{On the Schwinger effect during axion inflation}

\author[a]{Richard~von~Eckardstein,}
\author[a]{Kai~Schmitz,}
\author[a,b]{and Oleksandr~Sobol}

\affiliation[a]{Institute for Theoretical Physics, University of M\"unster,\\
Wilhelm-Klemm-Stra{\ss}e 9, 48149 M\"{u}nster, Germany}
\affiliation[b]{Physics Faculty, Taras Shevchenko National University of Kyiv,\\
64/13, Volodymyrska Street, 01601 Kyiv, Ukraine}

\emailAdd{richard.voneckardstein@uni-muenster.de}
\emailAdd{kai.schmitz@uni-muenster.de}
\emailAdd{oleksandr.sobol@uni-muenster.de}


\abstract{Pair-creation of charged particles in a strong gauge-field background\,---\, the renowned Schwinger effect\,---\, can strongly alter the efficiency of gauge-field production during axion inflation. It is therefore crucial to have a clear understanding and proper description of this phenomenon to obtain reliable predictions for the physical observables in this model. In the present work, we revisit the problem of Schwinger pair production during axion inflation in the presence of both electric and magnetic fields and improve on the state of the art in two ways: (i)~taking into account that the electric- and magnetic-field three-vectors are in general not collinear, we derive the vector decomposition of the Schwinger-induced current in terms of these fields and determine the corresponding effective electric and magnetic conductivities; (ii)~by identifying the physical momentum scale associated with the pair-creation process, we incorporate Schwinger damping of the gauge field in a scale-dependent fashion in the relevant equations of motion. Implementing this new description in the framework of the gradient-expansion formalism, we obtain numerical results in a benchmark scenario of axion inflation and perform a comprehensive comparison with earlier results in the literature. In some cases, the resulting energy densities of the produced gauge fields differ from the old results by more than one order of magnitude, which reflects the importance of taking the new effects into account.}


\maketitle


\section{Introduction}
\label{sec:introduction}


Cosmic inflation~\cite{Starobinsky:1980te,Guth:1980zm,Linde:1981mu,Starobinsky:1982ee,Albrecht:1982wi,Linde:1983gd} provides an elegant solution to the well-known puzzles of the Hot Big Bang model, such as the horizon, flatness, monopole problems, etc. Moreover, it can generate primordial inhomogeneities of the energy density and metric~\cite{Mukhanov:1981xt,Mukhanov:1982nu,Guth:1982ec,Hawking:1982cz,Bardeen:1983qw} which, later on, source the temperature anisotropies in the cosmic microwave background (CMB) and the large-scale structure of the Universe. Predictions of many slow-roll inflationary models for the properties of those inhomogeneities are in good agreement with observations~\cite{Planck:2018jri,BICEP:2021xfz,Reid:2009xm,Cabass:2022wjy}; see Refs.~\cite{Martin:2013tda,Chluba:2015bqa,Martin:2024qnn} for reviews. Another characteristic feature of inflation is that it generically produces tensor perturbations of the spacetime metric~\cite{Grishchuk:1974ny,Starobinsky:1979ty,Rubakov:1982df,Fabbri:1983us,Abbott:1984fp} which, being evolved until today, may manifest themselves as a stochastic gravitational-wave background.

However, all these astonishing features of inflation strongly rely on the flatness of the inflaton potential, which is hard to preserve in most particle-physics models. Indeed, the successful reheating of the Universe after inflation leading to a stage of radiation domination requires the inflaton field to couple to other matter fields in the Standard Model (SM)~\cite{Kofman:1997yn,Allahverdi:2010xz,Amin:2014eta}. On the other hand, these couplings typically give rise to large radiative corrections to the effective inflaton potential. These corrections violate its flatness, unless the latter is preserved by a certain symmetry. One of the renowned examples of this type is the axion inflation model, where the shift symmetry of the pseudoscalar inflaton field protects the potential from large radiative corrections. Originally, this model was proposed in Ref.~\cite{Freese:1990rb} under the name of natural inflation, which is now in tension with the CMB observations~\cite{Planck:2018jri,BICEP:2021xfz}; however, there are a few non-minimal realizations of axion inflation~\cite{Kim:2004rp,DAmico:2017cda} whose predictions lie within the observational bounds~\cite{Peloso:2015dsa,Copeland:2022lkp}.

More importantly, axion inflation admits a natural coupling of the pseudoscalar (axion) field $\phi$ to gauge fields via the term
\begin{equation}
    \mathcal{L}_{\mathrm{int}} = -\frac{\alpha_{\phi}}{4f}\,\phi\, F_{\mu\nu}\tilde{F}^{\mu\nu}\, ,
\end{equation}
which is also shift-symmetric in $\phi$. Here $\alpha_{\phi}$ is a dimensionless coupling constant, $f$ the characteristic scale arising from the axion's effective potential, $F_{\mu\nu}$ the gauge-field field-strength tensor, and $\tilde{F}^{\mu\nu}$ its dual. Note that these can be both Abelian and non-Abelian gauge fields. This interaction leads to the production of helical gauge fields during axion inflation resulting in a wide range of phenomenological applications. First, the produced gauge fields can be regarded as precursors of large-scale magnetic fields in the Universe~\cite{Turner:1987bw,Garretson:1992vt,Dolgov:1993vg,Anber:2006xt,Durrer:2010mq,Fujita:2015iga,Shtanov:2019civ,Shtanov:2019gpx,Sobol:2019xls}. They may source the baryon asymmetry of the Universe through the chiral anomaly~\cite{Anber:2015yca,Adshead:2016iae,Jimenez:2017cdr,Domcke:2019mnd,Domcke:2022kfs}. The presence of gauge fields during axion inflation may significantly change the dynamics of both the homogeneous background solution~\cite{Anber:2009ua,Adshead:2012kp,Notari:2016npn,Gorbar:2021rlt,Gorbar:2021zlr,Peloso:2022ovc,vonEckardstein:2023gwk,Durrer:2023rhc,Caravano:2021bfn,Caravano:2022epk,Domcke:2020zez,Figueroa:2023oxc,Domcke:2023tnn} and the perturbations on top of it~\cite{Barnaby:2010vf,Barnaby:2011vw,Bamba:2014vda,Ferreira:2014zia,Ferreira:2015omg,Cheng:2015oqa,Fujita:2015iga,Namba:2015gja,Dimastrogiovanni:2016fuu,Domcke:2016bkh,Papageorgiou:2018rfx,Papageorgiou:2019ecb,Domcke:2020zez,Dimastrogiovanni:2023oid,Domcke:2023tnn,Caravano:2022epk,Figueroa:2023oxc,Durrer:2024ibi,Sharma:2024nfu} leading to the formation of primordial black holes~\cite{Linde:2012bt,Bugaev:2013fya,Garcia-Bellido:2016dkw,Garcia-Bellido:2017aan,Ozsoy:2023ryl}, non-Gaussianities and parity violation in the CMB~\cite{Barnaby:2011qe,Barnaby:2012xt,Meerburg:2012id,Sorbo:2011rz}, and potentially observable features in the stochastic gravitational-wave background~\cite{Cook:2011hg,Garcia-Bellido:2016dkw,Domcke:2016bkh,Thorne:2017jft,Garcia-Bellido:2017aan,Garcia-Bellido:2023ser,Corba:2024tfz}. Moreover, the axion--vector coupling provides an effective reheating mechanism that transfers the inflaton energy density to radiation during the first few oscillations around the minimum of the inflaton potential~\cite{Adshead:2015pva,Figueroa:2017qmv,Adshead:2018doq,Cuissa:2018oiw,Adshead:2019lbr,Adshead:2019igv,Figueroa:2021yhd}.

Finally, the gauge fields generated during axion inflation may be strong enough to spontaneously create physical pairs of particles and antiparticles charged under the corresponding gauge group\,---\,the phenomenon known as the Schwinger effect~\cite{Sauter:1931zz,Heisenberg:1936nmg,Schwinger:1951nm}.%
\footnote{Sometimes, this effect is also referred to as the Sauter--Schwinger effect to acknowledge the fact that it was first predicted by Fritz Sauter in 1931~\cite{Sauter:1931zz}, 20 years before Julian Schwinger~\cite{Schwinger:1951nm}. However, for the sake of brevity, we will call it the Schwinger effect throughout this work.}
In the cosmological context, the Schwinger effect was first studied analytically in a simple case of a constant and homogeneous electric field in de Sitter spacetime in Refs.~\cite{Frob:2014zka,Kobayashi:2014zza,Bavarsad:2016cxh,Hayashinaka:2016dnt,Sharma:2017ivh,Hayashinaka:2018amz,Rajeev:2019okd} for scalar charge carriers and in Refs.~\cite{Stahl:2015gaa,Hayashinaka:2016qqn,Stahl:2016geq,Hayashinaka:2018amz} for charged fermions.%
\footnote{See also a recent work \cite{Kulkarni:2024kzc} studying direct production of fermions by a time-dependent axion field.}
Later on, in Refs.~\cite{Bavarsad:2017oyv,Domcke:2018eki,Domcke:2019qmm}, a constant and homogeneous magnetic field, (anti-)collinear to the electric one, was added. In the strong-field limit, $|\gc Q E|\gg H^2$ (where $E$ is the absolute value of the electric field, $\gc$ the gauge coupling constant, $Q$ the particle's charge, and $H$ the constant Hubble expansion rate of de Sitter spacetime), one recovers the corresponding expressions for flat Minkowski spacetime (for reviews, see Refs.~\cite{Dunne:2004nc,Ruffini:2009hg}) computed at the moment of time $t=1/(3H)$ after the field is turned on. The induced current of created particles has the form
\begin{equation}
     J = \frac{\big(\gc |Q|\big)^3}{6\pi^2 H} E |B| \, \psi\!\left( \frac{\pi |B|}{E}\right)e^{-\frac{\pi m^2}{\gc |Q| E}}\, ,
    \label{eq: current-colinear-frame}
\end{equation}
where $E=|\boldsymbol{E}|$, $J$ and $B$ are the projections of the electric current and magnetic field on the direction of the electric field, $m$ is the particle's mass and the function $\psi(x)=\operatorname{coth}x$ for the case of one Dirac fermion species and $\psi(x) = 1/(2\sinh x)$ for one complex scalar species. In the opposite case of a weak electric field, $|\gc Q E|\ll H^2$, the behavior of the Schwinger current was found to be significantly different from the Minkowski case and even somewhat counter-intuitive, exhibiting an infrared hyperconductivity for light scalars~\cite{Kobayashi:2014zza,Hayashinaka:2016dnt,Bavarsad:2017oyv} and negative conductivity for fermions~\cite{Hayashinaka:2016qqn,Hayashinaka:2018amz}. However, in Ref.~\cite{Banyeres:2018aax}, it was argued that these effects are spurious and connected to the running of the gauge coupling and higher-order corrections to the gauge-field action. Moreover, their impact on the gauge-field dynamics is negligibly small compared to the Hubble friction~\cite{Sharma:2017eps}. Therefore, in applications to realistic models of inflation, only the large-field expressions are typically considered ~\cite{Tangarife:2017rgl,Sobol:2018djj,Sobol:2019xls,Shtanov:2020gjp,Gorbar:2021rlt,Gorbar:2021zlr,Gorbar:2021ajq,Domcke:2021yuz,Cado:2022pxk,Fujita:2022fwc,Bastero-Gil:2023htv}.

Before applying the above expression for the Schwinger-induced current to real, time-dependent problems, one should address several issues. Most important among these issues is that the local-in-time relation between the induced current and the gauge field in Eq.~\eqref{eq: current-colinear-frame} may no longer be valid in a fully dynamical gauge-field background. Indeed, due to the inertial properties of the charge carriers, the Schwinger current may have a retarded response to changes in the gauge field. To track this behavior, one has to determine the current dynamically, on the same footing with the gauge field, e.g., within the hydrodynamic or kinetic framework~\cite{Gorbar:2019fpj,Sobol:2020frh,Lysenko:2023wrs,Gorbar:2023zla}. However, these studies show that, during slow-roll inflation, retardation effects are typically small, and one may still use the local-in-time form of the Schwinger current. The same conclusion was reached in Ref.~\cite{Kitamoto:2018htg}, where the induced current was computed in the Wentzel--Kramers--Brillouin approximation.

The most widely used and convenient representation of the Schwinger current is in the form of a generalized Ohm's law with the conductivity itself depending on the gauge-field configuration. With this Ohmic description, which we also adopt in the present paper, it is easy to incorporate the Schwinger effect in the equations of motion for the gauge field both in Fourier and position space~\cite{Domcke:2018eki,Domcke:2019mnd,Domcke:2019qmm,Gorbar:2021rlt,Gorbar:2021zlr}. In the case of constant and (anti-)collinear electric and magnetic fields, the current is proportional to both $E$ and $B$, see Eq.~\eqref{eq: current-colinear-frame}. Therefore, three possibilities present themselves for writing the generalized Ohm's law, e.g.,
\begin{subequations}
\begin{align}
   \bm{J}&=\sigma_{E} \bm{E} \quad \text{with}\quad  \sigma_E = \frac{\big(\gc |Q|\big)^3}{6\pi^2 H} |B| \, \psi\!\left( \frac{\pi |B|}{E}\right)e^{-\frac{\pi m^2}{\gc |Q| E}}\,  \label{eq: electric picture}\\
\intertext{or}
    \bm{J}&=\sigma_{B} \bm{B} \quad \text{with}\quad \sigma_{B} = \frac{\big(\gc |Q|\big)^3}{6\pi^2 H} \operatorname{sign}(B) E\,  \psi\!\left( \frac{\pi |B|}{E}\right)e^{-\frac{\pi m^2}{\gc |Q| E}}\, \label{eq: magnetic picture}\\
\intertext{or even in a more general form}
    \bm{J}&=\sigma_E \bm{E} + \sigma_{B} \bm{B}\, \label{eq: mixed picture}
\end{align}
\label{eq: pictures}%
\end{subequations}
with an arbitrary choice of conductivities $\sigma_{E/B}$ satisfying Eq.~\eqref{eq: current-colinear-frame}.
All three possibilities are equivalent at the classical level but lead to different pictures in the quantum case, where it makes a difference which quantities are treated as classical background quantities and which are promoted to quantum operators. Typically, the Schwinger conductivities are assumed to be classical functions depending only on time (i.e., on the time-dependent mean gauge field); therefore, the Schwinger current becomes linear in the gauge-field operators. The conventions $\hat{J}=\sigma_E \hat{E}$, $\hat{J}=\sigma_B \hat{B}$, and the combination of the two leads to different physical consequences.%
\footnote{Note that, in principle, one could consider a more general relation between the gauge field and current, $J^{i}=\sigma^{ij}_{E}E^{j}+\sigma^{ij}_{B}B^{j}$, where the electric and magnetic conductivities are second-rank three-tensors. However, when the conductivities are computed on the mean-field background, statistical isotropy of the inflationary background implies that these tensors should be reduced to the scalar functions multiplied by the Kronecker symbol $\delta^{ij}$ and no off-diagonal components are expected to appear. Therefore, we do not analyze this possibility in the present work.}
For definiteness, we will refer to the quantum versions of Eqs.~\eqref{eq: electric picture}, \eqref{eq: magnetic picture}, and \eqref{eq: mixed picture} as the electric, the magnetic, and the mixed picture, respectively. Note that the electric and magnetic pictures have already been discussed in the literature: the authors of Refs.~\cite{Domcke:2018eki,Domcke:2019mnd,Domcke:2019qmm} employed the magnetic picture to derive an upper bound on the maximal gauge-field strength that can be reached during axion inflation, while the authors of Refs.~\cite{Gorbar:2021rlt,Gorbar:2021zlr} used the electric picture to study gauge-field production in the language of the gradient-expansion formalism (GEF). In the present work, we shall resolve this ambiguity by explicitly considering the case of non-collinear electric and magnetic fields, which reflects the actual situation in the axion inflation model.

To derive an expression for the Schwinger current at the classical level, we will employ the procedure typically used in flat Minkowski spacetime~\cite{Dunne:2004nc,Ruffini:2009hg} (also applied in the cosmological context in Ref.~\cite{Fujita:2022fwc,Gorbar:2023zla}): We perform a boost to the reference frame where the fields are collinear, use the expression in Eq.~\eqref{eq: current-colinear-frame} for the current, and then boost back to the original reference frame. In this way, we obtain the Schwinger current three-vector as a linear combination of three-vectors of the electric and magnetic fields,
\begin{equation}
    \bm{J} = \sigma_E \bm{E} + \sigma_{B} \bm{B}\, ,
    \label{eq: current-linear-combination}
\end{equation}
where the scalar functions $\sigma_{E/B}$ are unambiguously expressed in terms of gauge-field scalar products; see Eq.~\eqref{eq: sigma-E-B} below. Our idea is then to promote the three-vector quantities in Eq.~\eqref{eq: current-linear-combination} to quantum operators treating the conductivities $\sigma_{E/B}$ as classical.  Recently, the authors of Ref.~\cite{Gorbar:2023zla} similarly split the Schwinger-induced current into magnetic and electric contributions; however, this splitting was based on completely different physical considerations, namely, on the separation of contributions to the induced current from the lowest and higher Landau levels as it was done in Ref.~\cite{Domcke:2018eki}. For the Schwinger conductivities, they used expressions derived and valid only in the collinear frame.
We would like to stress that our approach, as well as the one in Ref.~\cite{Gorbar:2023zla}, is an effective description of the Schwinger-induced current where the separation between classical quantities and quantum operators is performed by hand and is not based on any first-principles computation in the quantum field theory. However, this is the best thing one can do within the Ohmic approximation for the induced current. Ultimately, our results will need to be validated by a proper first-principles computation in the future.

In addition to the question of the correct picture for the Schwinger-induced current (electric, magnetic, or mixed), the Schwinger backreaction of the conductive medium on the gauge-field evolution gives rise to a second conceptual problem. Naively, one could think of the conductivities $\sigma_{E/B}$ as classical functions of time that are globally defined for all Fourier modes of the gauge field. This assumption was used in Refs.~\cite{Gorbar:2021rlt,Gorbar:2021zlr,Cado:2022pxk}, where it led to strong damping of the gauge-field modes, including modes deep inside the horizon, and to a significantly non-Markovian time evolution of the physical system: the produced gauge fields at the end of inflation were found to depend on the whole dynamical history of inflation, starting from the infinite past. On the other hand, from a microscopic point of view, it is clear that a gauge-field mode ought not to feel the presence of the conducting medium if its wavelength is shorter than the characteristic length scale of Schwinger pair production. Also, since the Schwinger effect is a causal phenomenon, it cannot occur if its characteristic length scale is larger than the size of causally connected patches in the universe, i.e., the Hubble horizon. To the best of our knowledge, a description of the Schwinger effect addressing the above-mentioned conceptual issues is still missing in the literature\,---\, a shortcoming that we attempt to remedy in the present paper.

In this work, we shall implement the description of the Schwinger effect discussed above in the GEF~\cite{Gorbar:2021rlt} and employ this improved version of the GEF to describe the gauge-field and charged-fermion production during axion inflation. Since we now have two characteristic momentum scales in the problem (i.e., the tachyonic-instability scale for the gauge-field production and the Schwinger pair-production scale), we are led to work with what basically amounts to a double copy of the GEF system of equations. In each copy, the gradient expansion is performed with respect to its own characteristic momentum scale, while both systems are coupled to each other via the Schwinger conductivities. We will provide the new results obtained in the mixed picture with the scale-dependent Schwinger damping and compare them to the results of the existing approaches in the literature, i.e., the electric and magnetic pictures with no scale dependence in the Schwinger conductivities. Moreover, we will propose a simplified treatment for the scale dependence that is computationally less costly and yields a satisfactory agreement with the exact result.

The remainder of this paper is organized as follows. We will review the basics of the axion inflation model in section~\ref{sec:axion-inflation}. In section~\ref{sec:schwinger}, we will then present our improved description of the Schwinger effect, including the splitting of the Schwinger current into electric and magnetic components and the introduction of the scale (or momentum) dependence of the conductivities. We will incorporate this new description in the GEF in section~\ref{sec:GEF}. Section~\ref{sec:results} is devoted to numerical results in a benchmark scenario of axion inflation with a detailed comparison to the existing approaches in the literature. In section.~\ref{sec:conclusions}, we will present our conclusions. Appendix~\ref{app: Whittaker} contains information about the Whittaker functions and their properties. In appendix~\ref{app: rhoChi EoM}, we sketch the derivation of the equation of motion for the energy density of the charged particles produced via the Schwinger effect.

\smallskip
\noindent\textbf{Conventions:} throughout this work, we use natural units and set $\hbar=c=1$; we use the notation $M_{\mathrm{P}}=(8\pi G)^{-1/2}\approx \qty{2.43e18}{\giga\eV}$ for the reduced Planck mass; and we assume that the Universe is described by a spatially flat Friedmann--Lema\^{i}tre--Robertson--Walker (FLRW) metric with line element (in terms of cosmic time $t$ and conformal time $\eta$)
\begin{equation}
    ds^2 = g_{\mu \nu} \D x^\mu \D x^\nu = \D t^2 - a^2(t)\D \bm{x}^2 = a^2(\eta)\left[\D \eta^2 - \D \bm{x}^2 \right]\, .
    \label{eq: metric}
\end{equation}
An overdot denotes a derivative with respect to cosmic time; all other derivatives are written explicitly.


\section{Axion inflation}
\label{sec:axion-inflation}


We consider a model of axion inflation where the pseudoscalar inflaton field $\phi$ couples to a $U(1)$ gauge field $A_\mu$ via a Chern--Simons-like term,
\begin{equation}
    S = \int \D^4 x \sqrt{-g} \left[\frac{1}{2} g^{\mu \nu} \p_{\mu} \phi \,  \p_{\nu} \phi - V(\phi) - \frac{1}{4} F_{\mu \nu} F^{\mu \nu} - \frac{1}{4} I(\phi) F_{\mu \nu} \tilde{F}^{\mu \nu} + \mathcal{L}_{\mathrm{ch}}(\chi, A_{\mu})\right]\, ,
    \label{eq: Action}
\end{equation}
where $V(\phi)$ is the inflaton potential, $F_{\mu \nu} = \p_\mu A_\nu - \p_\nu A_\mu$ the gauge-field field-strength tensor, $\tilde{F}^{\mu \nu} = \varepsilon^{\mu \nu \alpha \beta}/(2\sqrt{-g})\,F_{\alpha \beta}$ its dual, where $\varepsilon^{0123} = 1$, and $I(\phi)$ is a generic axial coupling function. The last term in Eq.~\eqref{eq: Action}, $\mathcal{L}_{\mathrm{ch}}$, is the gauge-invariant Lagrangian describing a generic matter field $\chi$ that is charged under the $U(1)$ gauge group and hence couples to the gauge field $A_\mu$. For the sake of generality, we do not specify this Lagrangian and simply assume that it describes all charged fields in the model.

From the action in Eq.~\eqref{eq: Action}, we can derive the Euler--Lagrange equations for the inflaton and gauge fields. Supplementing them with the Bianchi identity for the dual field-strength tensor $\tilde{F}^{\mu\nu}$ leads to the following set of equations of motion:
\begin{subequations}
    \begin{align}
        \frac{1}{\sqrt{-g}}\p_\alpha \left[\sqrt{-g} g^{\alpha \mu} \p_\mu \phi\right] + V_{,\phi} + \frac{1}{4} I_{,\phi} F_{\mu \nu} \tilde{F}^{\mu \nu} &= 0\, , \vphantom{\Bigg(} \label{eq: KG General}\\
        \frac{1}{\sqrt{-g}}\p_\alpha \left[\sqrt{-g} g^{\mu \alpha} g^{\nu \beta} F_{\mu \nu}\right] + I_{,\phi} \tilde{F}^{\alpha \beta} \p_\alpha \phi 
        &= J^\beta ,\vphantom{\Bigg(} \label{eq: Maxwell General}\\
        \frac{1}{\sqrt{-g}}\p_\alpha \left[\sqrt{-g} \tilde{F}^{\alpha \beta}\right] &= 0\, . \vphantom{\Bigg(} \label{eq: Bianchi General}
    \end{align}
    \label{eq: EoM General}%
\end{subequations}
Here, the lower index $_{,\phi}$ denotes the derivative w.r.t.\ the inflaton field, $X_{,\phi}\equiv dX/d\phi$, and we defined the electric four-current $J^\mu$ through
\begin{equation}
     J^\mu = - \frac{\p \mathcal{L}_{\mathrm{ch}}}{\p A_\mu}\, .
\end{equation}
Varying the action \eqref{eq: Action} with respect to the metric yields the energy--momentum tensor
\begin{equation}
    T_{\mu \nu} = \p_{\mu} \phi \, \p_{\nu} \phi - g^{\alpha \beta} F_{\mu \alpha} F_{\nu \beta} - g_{\mu \nu} \left[\frac{1}{2} \p_{\alpha} \phi \,  \p^{\alpha} \phi - V(\phi) - \frac{1}{4}  F_{\alpha \beta} F^{\alpha \beta}\right] +  T_{\mu \nu}^\chi\, ,
    \label{eq: General E-M-tensor}
\end{equation}
where the last term describes the contribution of charged matter fields.

We now assume the FLRW metric expressed in physical time $t$ and consider the inflaton field as a homogeneous background field $\phi(t, \bm{x}) = \phi(t)$. These assumptions allow us to choose temporal and Coulomb gauge%
\footnote{Note that the additional choice of temporal gauge is only valid if one assumes a homogeneous inflaton field as implied by the Maxwell equations for this model. Furthermore, it is noteworthy that the Schwinger effect produces no net electric charge. If $J_0 = 0$ initially, it will thus stay zero throughout inflation.}
for the vector potential $A_\mu$, such that $A_\mu = (0, -\bm{A})$. We then define the electric and magnetic%
\footnote{We stress that $A_\mu$ is a generic Abelian gauge field. However, throughout the paper, we often refer to it as the electromagnetic field and to its three-vectors as the electric and magnetic fields. This is done just by analogy with the SM electromagnetic field and does not mean that the analysis is restricted to this case.}
three-vector fields $\bm{E} = (E^1, E^2, E^3)$ and $\bm{B} = (B^1, B^2, B^3)$ through
\begin{subequations}
    \begin{equation}
        \bm{E}=-\frac{1}{a}\p_0\bm{A}\, , \qquad \bm{B}=\frac{1}{a^2}\mathrm{rot}\bm{A}\, ,
    \end{equation}
    \begin{equation}
        F_{0 i} = a E^i\, , \quad F_{i j}= - a^2 \varepsilon_{ijk} B^k\, , \quad \tilde{F}_{0 i} = a B^i\, , \quad  \tilde{F}_{i j} = a^2 \varepsilon_{ijk} E^k\, .
    \end{equation}
    \label{eq: EandBthroughF}%
\end{subequations}
Note that these are the physical fields that a comoving observer would measure, which is why we include the scale factor in their definition.

The Friedmann equation determines the evolution of the scale factor through the Hubble rate $H = \dot{a}/a$,
\begin{equation}
    H^2 = \frac{\rho}{3 M_{\mathrm{P}}^2}\, ,
    \label{eq: Friedmann}
\end{equation}
with the total energy density $\rho$ given by the zero--zero component of the energy--momentum tensor \eqref{eq: General E-M-tensor},
\begin{equation}
    \rho = T_0^0 = \bigg[\frac{1}{2}\dot{\phi}^2 + V(\phi)\bigg]  + \frac{1}{2}(\langle\bm{E}^2\rangle + \langle\bm{B}^2\rangle) + \rho_\chi\, .
    \label{eq: rho}
\end{equation}
The first two terms in square brackets describe the energy density of the homogeneous inflaton field. The subsequent term gives the gauge-field contribution to the total energy density, and $\rho_\chi$ is the energy density of the charged matter fields.

Further, we decompose the electric four-current as
\begin{equation}
    J^\mu = (\rho_{\mathrm{ch}}, \frac{1}{a}\bm{J})\, .
\end{equation}
Assuming that charged particles were initially absent in the Universe, and only produced later in particle--antiparticle pairs, allows us to set the charge density to zero, $\rho_{\mathrm{ch}} = 0$. However, the electromagnetic fields may induce a three-vector charge current, $\bm{J}$. Then, the equations of motion \eqref{eq: EoM General} take the following form in three-vector notation:
\begin{subequations}
    \begin{equation}
        \Ddot{\phi} + 3H\dot{\phi} + V_{,\phi} = \frac{1}{2} I_{,\phi} \braket{\bm{E} \cdot \bm{B} + \bm{B} \cdot \bm{E}}\, ,
        \label{eq: phiEoM}
    \end{equation}
    \begin{equation}
        \diver \bm{E} = 0\, , \qquad \diver \bm{B} = 0\, ,
    \end{equation}
    \begin{equation}
        \dot{\bm{E}} + 2 H \bm{E} - \frac{1}{a}\rot \bm{B} + I_{,\phi} \dot{\phi}\bm{B} + \bm{J} = 0\, ,
        \label{eq: EEoM}
    \end{equation}
    \begin{equation}
        \dot{\bm{B}}  + 2 H \bm{B} + \frac{1}{a}\rot \bm{E} = 0\, .
        \label{eq: BEoM}
    \end{equation}
    \label{eq: EoMs}%
\end{subequations}%
These are the governing equations of motion for axion inflation in an expanding background, described by the Friedmann equation \eqref{eq: Friedmann}, in the presence of an electric current $\bm{J}$. Indeed, setting $\bm{J} = 0$ and $\rho_\chi = 0$ reduces these equations to the widely studied case of axion inflation without charged fermion production. To close this set of equations, we need to specify the electric current $\bm{J}$, which is the topic of the following section.


\section{Schwinger effect during axion inflation}
\label{sec:schwinger}


\subsection{Electric and magnetic conductivities}
\label{subsec: Sigmas}

As discussed in the introduction, an analytical expression for the Schwinger-induced current in an expanding background is known only for the simplest case of constant and spatially uniform (anti-)collinear electric and magnetic fields in de Sitter spacetime. In the strong-field limit, $|\gc QE|\gg H^2$, it is given by Eq.~\eqref{eq: current-colinear-frame}~\cite{Bavarsad:2017oyv,Domcke:2018eki,Domcke:2019qmm}. Since, in this frame, the vectors of the electric field, magnetic field, and induced current are all collinear, there is an ambiguity in writing Ohm's law for the Schwinger current: one may choose to represent it in terms of the conventional electric conductivity, $\bm{J} = \sigma_E \bm{E}$, as in Refs.~\cite{Kobayashi:2014zza, Hayashinaka:2016qqn}, or consider the magnetic picture where $\bm{J} = \sigma_B \bm{B}$ as in Refs.~\cite{Domcke:2018eki,Domcke:2019qmm}. In what follows, we shall address this ambiguity in the description of the current by dropping the assumption on the collinearity of the electric and magnetic fields.

Numerical solutions to the equations of motion of axion inflation in the presence of electric conductivity demonstrate that the electric and magnetic fields are only approximately collinear in the comoving frame \cite{Gorbar:2021rlt}. However, one can always perform a Lorentz boost to a coordinate system in which the two fields are collinear and where the expression \eqref{eq: current-colinear-frame} is valid. It was shown in Ref.~\cite{Fujita:2022fwc} that boosting from the comoving coordinate frame to a frame where the electromagnetic fields are collinear, evaluating Eq.~\eqref{eq: current-colinear-frame}, and boosting back to the comoving frame lifts the dual description of the induced current in terms of Ohm's law. This procedure requires expressing the induced current through both an electric and a magnetic conductivity,
\begin{equation}
    \bm{J} = \sigma_E \bm{E} + \sigma_B \bm{B}\, .
    \label{eq: current Ansatz}
\end{equation}
While the authors of Ref.~\cite{Fujita:2022fwc} studied small perturbations around constant, anticollinear electric and magnetic background fields, we will go beyond perturbation theory and derive fully general expressions for the electric and magnetic conductivities, applying the same technique as in Ref.~\cite{Fujita:2022fwc}: Lorentz-boosting to the collinear frame and back again.

To boost to the collinear frame, we only need to know the angle between $\bm{E}$ and $\bm{B}$ in the comoving frame. A boost with velocity $\bm{v}$ and Lorentz factor $\gamma=(1-v^2)^{-1/2}$ transforms the fields as follows (see, e.g., the textbook \cite{Maggiore:2023ClassElectrod}):
\begin{equation}
    \bm{E}'=\gamma(\bm{E}+\bm{v}\times\bm{B})-\frac{\gamma^2}{1+\gamma}\bm{v}(\bm{v}\cdot\bm{E})\, ,\qquad \bm{B}'=\gamma(\bm{B}-\bm{v}\times\bm{E})-\frac{\gamma^2}{1+\gamma}\bm{v}(\bm{v}\cdot\bm{B})\, .
\end{equation}
Requiring that in the collinear frame $\bm{E}'\times\bm{B}'=0$ and choosing from the infinitely many possibilities the velocity which is perpendicular to both the electric and magnetic field, i.e., $\bm{v}=\kappa \,\bm{E}\times\bm{B}$, we immediately get a simple quadratic equation for the coefficient $\kappa$:
\begin{equation}
    \kappa^2 |\bm{E}\times\bm{B}|^2 - \kappa (\bm{E}^2 + \bm{B}^2) + 1=0\, .
\end{equation}
Its solution implies the boost velocity which rotates the fields into an (anti-)collinear configuration in the form:
\begin{equation}
    \bm{v} = \kappa \,\bm{E}\times\bm{B}\,, \qquad \kappa= \frac{2}{\bm{E}^2 + \bm{B}^2 + \sqrt{\left(\bm{E}^2 - \bm{B}^2\right)^2 + 4\left(\bm{E}\cdot\bm{B}\right)^2}}\, .
    \label{eq: Boost velocity}
\end{equation}
The corresponding Lorentz factor is given by
\begin{equation}
    \gamma = \frac{1}{\sqrt{1-v^2}} = \frac{1}{\sqrt{2}}\left[1 + \frac{\bm{E}^2 + \bm{B}^2}{\sqrt{\left(\bm{E}^2 - \bm{B}^2\right)^2 + 4\left(\bm{E}\cdot\bm{B}\right)^2}} \right]^{1/2}\, .
    \label{eq: Lorentz factor}
\end{equation}
We can determine the values of the electric and magnetic fields in the collinear frame from the (frame-independent) invariants of the field-strength tensor
\begin{subequations}
    \begin{align}
        \frac{1}{2} F_{\mu \nu}F^{\mu \nu} &= \bm{B}^2 -\bm{E}^2 = {B'}^2 - {E'}^2\, ,\\
        -\frac{1}{4} F_{\mu \nu}\tilde{F}^{\mu \nu} &= \bm{E} \cdot \bm{B} = E' B'\, ,
    \end{align}
\end{subequations}
where $E'$ and $B'$ denote the magnitudes of the electric and magnetic fields in the collinear frame. From these equations, one can immediately derive that
\begin{subequations}
    \begin{align}
        E' &= \frac{1}{\sqrt{2}} \left[\bm{E}^2 -\bm{B}^2 + \sqrt{\left(\bm{E}^2 - \bm{B}^2\right)^2 + 4\left(\bm{E}\cdot\bm{B}\right)^2}\right]^{1/2}\, ,\label{eq: Collinear E-prime}\\
        B' &= \operatorname{sign} (\bm{E}\cdot\bm{B})\frac{1}{\sqrt{2}} \left[\bm{B}^2 -\bm{E}^2 + \sqrt{\left(\bm{E}^2 - \bm{B}^2\right)^2 + 4\left(\bm{E}\cdot\bm{B}\right)^2}\right]^{1/2}\, .
    \end{align}
    \label{eq: Collinear through Comoving EB}%
\end{subequations}%
Note that we define $E'$ as strictly positive, while $B'$ is the projection of the magnetic field onto the electric-field direction, which can be negative in the anticollinear case.

Since the current $\bm{J} = \sigma_E \bm{E} + \sigma_B \bm{B}$ is orthogonal to the velocity in Eq.~\eqref{eq: Boost velocity}, the boost will leave it invariant, $\bm{J} = \bm{J}'$. In the collinear frame, $\bm{J'}$ can be directly taken from Eq.~\eqref{eq: current-colinear-frame} as this expression was derived assuming collinear fields:
\begin{equation}
    |\bm{J}'| = \frac{\big(\gc |Q|\big)^3}{6\pi^2 H} E' |B'| \, \psi\!\left( \frac{\pi |B'|}{E'}\right)e^{-\frac{\pi m^2}{\gc |Q| E'}}\, .
    \label{eq: current in collinear frame 2}
\end{equation}
We see that the current in the collinear frame is collinear with the electric field $\bm{E}'$ allowing us to write down the following relation
\begin{align}
    \bm{J} = |\bm{J}'|\frac{\bm{E'}}{E'}&= \frac{|\bm{J}'|\gamma}{E'}\left[\bm{E} +\kappa\,\bm{B} \times (\bm{B} \times \bm{E})\right]\nonumber\\
    &=\frac{|\bm{J}'|\gamma}{E'}\left[(1-\kappa \bm{B}^2)\bm{E} + \kappa(\bm{E}\cdot\bm{B})\bm{B}\right]= \sigma_E \bm{E} + \sigma_B \bm{B}\, 
    \label{eq: Current Decomposed into E and B fields}
\end{align}
with
\begin{subequations}
    \begin{align}
        \sigma_E &= \frac{|\bm{J}'| \gamma}{E'} (1-\kappa \bm{B}^2)
        = \frac{|\bm{J}'| E'}{\gamma} \frac{1}{\sqrt{\left(\bm{E}^2 - \bm{B}^2\right)^2 + 4\left(\bm{E}\cdot\bm{B}\right)^2}}\, ,\\
        \sigma_B &= \frac{|\bm{J}'| \gamma}{E'}\kappa(\bm{E}\cdot\bm{B}) 
        = \frac{|\bm{J}'|}{E' \gamma} \frac{(\bm{E} \cdot \bm{B})}{\sqrt{\left(\bm{E}^2 - \bm{B}^2\right)^2 + 4\left(\bm{E}\cdot\bm{B}\right)^2}}\, ,
    \end{align}
    \label{eq: sigma-E-B}%
\end{subequations}
where we used Eqs.~\eqref{eq: Boost velocity}, \eqref{eq: Lorentz factor}, and \eqref{eq: Collinear E-prime} in the second line of each respective equation. Using Eq.~\eqref{eq: Lorentz factor}, Eqs.~\eqref{eq: Collinear through Comoving EB}, and Eq.~\eqref{eq: current in collinear frame 2} to evaluate $\gamma$, $E'$, $B'$ and $|\bm{J}'|$, we can finally express $\sigma_E$ and $\sigma_B$ in Eqs.~\eqref{eq: sigma-E-B} entirely in terms of $|\bm{E}|$, $|\bm{B}|$, and $(\bm{E} \cdot \bm{B})$.

Taking the results of this section, we can return to Amp{\`e}re's law in an expanding background with a charged current and an axion interaction term, Eq.~\eqref{eq: EEoM}. Using our expression for the induced current $\bm{J}$, Eq.~\eqref{eq: current Ansatz}, we can express Amp{\`e}re's law as
\begin{equation}
    \dot{\bm{E}} + (2 H + \sigma_E) \bm{E} - \frac{1}{a}\rot \bm{B} + \left(I_{,\phi} \dot{\phi} + \sigma_B \right)\! \bm{B} = 0\, .
    \label{eq: EEoM with conductivities}
\end{equation}
With the expressions for $\sigma_E$ and $\sigma_B$, Amp{\`e}re's law is now entirely expressed in terms of $\bm{E}$, $\bm{B}$, and $(\bm{E} \cdot \bm{B})$ as well as the inflaton velocity and the Hubble rate.
Note that both conductivities typically lead to the damping of the gauge field: the (positive) electric conductivity adds up to the Hubble friction while the magnetic one suppresses the gauge-field production by decreasing the term with the axial coupling because $\sigma_{B}\propto (\bm{E}\cdot\bm{B})$ which typically has the opposite sign to $I_{,\phi}\dot{\phi}$.


\subsection{Damping in a conductive medium}
\label{subsec: Delta}

Let us take into account the effect of a conductive medium produced by the Schwinger effect on the evolution of the gauge-field modes. To this end, consider the quantized gauge field
\begin{equation}
     \hat{\bm{A}}(t, \bm{x}) = \int \frac{\D ^3 \bm{k}}{(2 \pi)^{3/2}} \sum_{\lambda = \pm} \left[\bm{\epsilon}^\lambda(\bm{k}) \hat{a}^{\vphantom{\dagger}}_{\bm{k}, \lambda} A_\lambda(t,k) e^{i \bm{k} \cdot \bm{x}} + \bm{\epsilon}^{\ast\lambda}(\bm{k}) \hat{a}_{\bm{k}, \lambda}^\dagger A^{\ast}_\lambda(t,k) e^{-i \bm{k} \cdot \bm{x}}\right]\, ,
     \label{eq: FourierModes}
\end{equation}
where $A_\lambda(t,k)$ is the mode function at time $t$ with circular polarization $\lambda = \pm$, and wavenumber $k = |\bm{k}|$. The polarization three-vectors, $\bm{\epsilon}^\lambda(\bm{k})$, satisfy
\begin{equation}
\arraycolsep=10pt
    \begin{array}{ll}
         \bm{k} \cdot \bm{\epsilon}^\lambda(\bm{k}) = 0\, , \quad &  i\bm{k}\times \bm{\epsilon}^\lambda(\bm{k}) = \lambda k \bm{\epsilon}^\lambda(\bm{k})\, ,\\[8pt]
         \bm{\epsilon}^\lambda(\bm{k}) \cdot \bm{\epsilon}^{\ast\lambda'}(\bm{k}) = \delta_{\lambda\lambda'}\, , \qquad &  \bm{\epsilon}^{\ast\lambda}(\bm{k}) = \bm{\epsilon}^{-\lambda}(\bm{k})= \bm{\epsilon}^{\lambda}(-\bm{k})\, .
    \end{array}
\end{equation}
The annihilation operators, $\hat{a}^{\vphantom{\dagger}}_{\bm{k}, \lambda}$, and creation operators, $\hat{a}^{\dagger}_{\bm{k},\lambda}$, of the electromagnetic mode with wavevector $\bm{k}$ and polarization $\lambda$ satisfy the canonical commutation relations,
\begin{equation}
    \big[\hat{a}^{\vphantom{\dagger}}_{\bm{k}, \lambda}, \hat{a}_{\bm{k}', \lambda'}^\dagger\big] = \delta_{\lambda\lambda'} \delta^{(3)}(\boldsymbol{k} - \boldsymbol{k'})\, .
\end{equation}
We can then deduce the equation of motion for the mode functions from Eq.~\eqref{eq: EEoM with conductivities},
\begin{equation}
    \ddot{A}_\lambda(t,k) + \big(H + \sigma_E \big)\dot{A}_\lambda(t,k) + \bigg[\bigg(\frac{k}{a}\bigg)^{\!\!2} - \lambda\, \frac{k}{a} \big(I_{,\phi} \dot{\phi} + \sigma_B \big)\bigg] A_\lambda(t,k) = 0\, .
    \label{eq: Mode Equation}
\end{equation}
To derive this equation, one has to neglect that the conductivities $\sigma_E$ and $\sigma_B$ depend themselves on the mode functions $A_\lambda$, as evident from Eq.~\eqref{eq: sigma-E-B}. A common approach in the literature to obtain the evolution of the gauge-field mode functions is to treat $\sigma_E$ and $\sigma_B$ as given functions of time determined from the classical electric and magnetic field values. In this work, we want to keep to this simplifying assumption, but we will additionally allow for a non-trivial dependence of the conductivities on the wavenumber $k$ of the specific mode under consideration. The reason for this should be intuitively clear: Gauge-field modes whose wavelength is much smaller than the typical separation between the fermions produced through Schwinger pair production will not notice their presence and, thus, the corresponding mode functions should not experience any damping due to the conductivity of the fermion plasma. We realize this idea by replacing all dependence on the functions $\sigma_E(t)$ and $\sigma_B(t)$ through
\begin{equation}
    \sigma_{E/B}(t,k)=\sigma_{E/B}(t) \Theta(t, k)\, ,
\end{equation}
where $\Theta(t, k)$ is some function capturing the scale dependence of Schwinger damping as the system evolves in time. Of course, this is an \textit{ad hoc} approach to describe the scale dependence in the model under investigation. To find this dependence, one should, in principle, consider the Schwinger pair-production process by a constant background gauge field in the presence of a high-frequency mode from first principles. However, this task is beyond the scope of the present article and deserves a separate investigation.

Introducing the proposed scale dependence for the conductivities by hand, the equation of motion for the gauge-field modes now reads
\begin{equation}
    \ddot{A}_\lambda(t,k) + \big(H + \sigma_E \Theta \big)\dot{A}_\lambda(t,k) + \bigg[\bigg(\frac{k}{a}\bigg)^{\!\!2} - \lambda\, \frac{k}{a} \big(I_{,\phi} \dot{\phi} + \sigma_B \Theta \big)\bigg] A_\lambda(t,k) = 0\, .
    \label{eq: Mode Equation, K-dep Sigmas}
\end{equation}
The equation is simplified by the Ansatz
\begin{equation}
    A_\lambda(t,k) = \sqrt{\Delta(t, k)}\: f_\lambda(t,k) = \exp\Bigg(-\frac{1}{2}\!\int\limits_{-\infty}^t \!\! \sigma_E(t') \Theta(t',k) \D t'\Bigg)f_\lambda(t,k)\, .
    \label{eq: Damped gauge-modes}
\end{equation}
It has a clear physical interpretation: The true mode functions $A_\lambda(t,k)$ are exponentially suppressed with respect to $f_\lambda(t,k)$ by the damping factor, $\Delta(t,k)$. This factor accounts for the damping of the gauge-field modes due to the conductive medium of the fermions and can also be understood as the additional accumulating friction due to the term $\sigma_E \dot{A_\lambda}(t,k)$ in Eq.~\eqref{eq: Mode Equation, K-dep Sigmas}. If one furthermore applies a change of variable from $t$ to $z = k \eta(t)$, with conformal time $\eta(t) = \int^t \D t' / a(t')$, one finds that the $f_\lambda$'s obey
\begin{equation}
    \frac{\D^2}{\D z^2}f_\lambda(z,k) + \bigg\{1 - \lambda\, \frac{a}{k} \big(I_{,\phi} \dot{\phi} + \sigma_B \Theta \big) - \frac{1}{2}\Big(\frac{a}{k}\Big)^{\!2} \Big[(\sigma_E \Theta)^{\bm{\cdot}} + \frac{1}{2}(\sigma_E \Theta)^2 + H \sigma_E \Theta \Big]\bigg\}f_\lambda(z,k) = 0\, .
\end{equation}
During inflation, we can approximate the cosmic expansion as nearly de Sitter, $\eta \approx -1/(aH)$. This gives
\begin{equation}
    \frac{\D^2}{\D z^2}f_\lambda(z,k) + \bigg[1 + \frac{2\lambda \xi_{\mathrm{eff}}}{z} - \frac{s_E+s_E^2}{z^2} \bigg]f_\lambda(z,k) = 0\, ,
    \label{eq: f-modes EoM}
\end{equation}
where we introduced the following useful variables,
\begin{equation}
    \xi = I_{,\phi}\frac{ \dot{\phi}}{2H}\, , \qquad \xi_{\mathrm{eff}} = \xi + \frac{\sigma_B}{2H} \Theta(t,k)\, , \qquad s_E = \frac{\sigma_E}{2H} \Theta(t,k)\, ,
    \label{eq: gauge-field production parameters}
\end{equation}
and assumed that $\dot{s}_E \approx 0$. This is true if the Hubble rate and the gauge-field evolve sufficiently slowly during inflation.

Let us examine Eq.~\eqref{eq: f-modes EoM} more carefully to gain some intuition about the various ingredients involved.
Firstly, we can see that deep inside the horizon, i.e., when $-z \to \infty$, the equation is approximately that of a harmonic oscillator. Imposing the Bunch--Davies boundary condition
\begin{equation}
    f_\lambda(z,k) = \frac{1}{\sqrt{2k}} e^{-iz}\, , \qquad A_\lambda(z,k) = \sqrt{\frac{\Delta(z,k)}{2k}} e^{-iz}\, , \qquad -z \gg 1\, ,
    \label{eq: BD}
\end{equation}
we ensure that the mode function $f_\lambda(z,k)$ contains only the positive-frequency solution that physically corresponds to the absence of gauge-field quanta with arbitrarily large momenta.

From this equation, we also easily recognize the well-known parity-violating tachyonic instability that the gauge modes $A_\lambda$ experience during axion inflation. The instability occurs for modes with wavenumber $k$ lower than the critical value
\begin{equation}
    \frac{k_{\mathrm{crit}}^\lambda}{a H} = \lambda \xi_{\mathrm{eff}} + \sqrt{\xi_{\mathrm{eff}}^2 + s_E^2 + s_E}\, .
    \label{eq: instability scale}
\end{equation}
Interestingly, in axion inflation without Schwinger pair production ($s_E = 0$, $\xi_{\mathrm{eff}}=\xi$), only the modes of one helicity, $\lambda=\operatorname{sign}\xi$, experience the instability and are exponentially amplified. However, in the presence of electric conductivity, $s_E > 0$, both helicity modes can be amplified. For the helicity with $\lambda=-\operatorname{sign}\xi_{\mathrm{eff}}$, though, the instability is experienced by fewer modes due to the lower critical threshold value $k_{\mathrm{crit}}^\lambda$. Note that this enhancement of modes with $\lambda=-\operatorname{sign}\xi_{\mathrm{eff}}$ occurs only for the rescaled mode function $f_\lambda$; the full mode function $A_\lambda$ does not experience any enhancement because of the extra exponential damping factor $\sqrt{\Delta}$ in Eq.~\eqref{eq: Damped gauge-modes}.

Lastly, assuming constant values for $\xi_{\mathrm{eff}}$ and $s_E$, Eq.~\eqref{eq: f-modes EoM} can be solved exactly in terms of the Whittaker functions $W_{\kappa, \mu}$ and $M_{\kappa, \mu}$. General properties of these functions can be found in appendix~\ref{app: Whittaker}. The mode functions $A_\lambda$ can then be written in terms of the two linearly independent solutions
\begin{equation}
    A_{\lambda}(\eta, k)=\sqrt{\frac{\Delta(\eta, k)}{2k}} e^{\frac{\lambda \pi \xi}{2}} \left[C_{\lambda}(k) W_{-i\lambda\xi_{\mathrm{eff}},\frac{1}{2}+s_E}(2ik\eta)+D_{\lambda}(k) M_{-i\lambda\xi_{\mathrm{eff}},\frac{1}{2}+s_E}(2ik\eta)\right]\, .
     \label{eq: general A mode solution}
\end{equation}
For this conclusion, we need to consider that $\xi_{\mathrm{eff}} = \mathrm{const}$, $s_E = \mathrm{const}$ and $\eta \approx -1/(aH)$, as explained above. Let us now consider two scenarios for treating the scale dependence of the electric and magnetic conductivity terms, i.e., the choice of the function $\Theta(t,k)$.

The most common choice in the literature so far is to simply set $\Theta(t,k)=1$; see, e.g., Ref.~\cite{Gorbar:2021rlt}. From Eq.~\eqref{eq: Damped gauge-modes}, we see that this implies
\begin{equation}
    \Delta(t,k) =  \Delta(t) = \exp \Bigg(-\int\limits_{-\infty}^t \sigma_E(t') \D t'\Bigg).
\end{equation}
In this scenario, the boundary condition in Eq.~\eqref{eq: BD} implies that the mode functions $A_\lambda(t,k)$ are approximately given by
\begin{equation}
    \label{eq: solution-no-scale-dep}
    A_{\lambda}(\eta, k)=\sqrt{\frac{\Delta(\eta)}{2k}} e^{\frac{\lambda \pi \xi}{2}} W_{-i\lambda\xi_{\mathrm{eff}}, \frac{1}{2}+s_E}(2ik\eta)\, .
\end{equation}
Evidently, all modes $A_\lambda$ will feel the presence of the conductivity induced by the fermions at all times. Even modes infinitely deep inside the horizon would feel the presence of the conductive medium, see Eq.~\eqref{eq: BD}.

This picture appears counter-intuitive given the physical nature of Schwinger pair production. Indeed, the creation of a particle--antiparticle pair occurs when the electric field is strong enough to make the virtual pair, which is always present in the physical vacuum, become a real pair. To estimate the characteristic time and spatial scale of pair creation, let us switch to the collinear frame and consider the case when the particle's mass is negligibly small, $m^2\ll |\gc QE'|$. Then, a particle with a small initial longitudinal projection of its momentum will move in the direction of the electric field, while the magnetic field will not impact its motion along this direction. On the other hand, the electric field will not change the motion in the transverse plane, i.e., the number of the Landau level occupied by the particle. In this way, the effects of the electric and magnetic field on the particle motion are disentangled in the collinear frame. To prevent annihilation, the electric field must separate the created particles and antiparticles in the longitudinal direction by a distance greater than the corresponding de Broglie wavelength, i.e., 
\begin{equation}
    \delta x' > \lambda_{\mathrm{dB}}=\frac{2\pi}{p}=\frac{2\pi}{|\gc QE'|t'}\simeq \frac{2\pi}{|\gc Q E'|\delta x'}\, ,
\end{equation}
where in the last step we took into account that the created particles are ultrarelativistic.

From this, we immediately obtain an estimate for the characteristic physical spatial and time scale for the Schwinger pair production in the collinear frame:
\begin{equation}
    \label{eq: Schwinger-timescale}
    t'_{\mathrm{S}}\sim \delta x'_{\mathrm{S}}\sim \frac{1}{\sqrt{|\gc Q E'|}}\, .
\end{equation}
Now, we can boost back to the comoving reference frame and take into account that $\delta x'_{\mathrm{S}}$ is oriented along the direction of the electric field $\bm{E}'$, i.e., perpendicular to the boost velocity. Then, the characteristic spatial scale in the comoving frame remains the same as in the collinear frame,
\begin{equation}
    \delta x_{\mathrm{S}}=\delta x'_{\mathrm{S}}\sim \frac{1}{\sqrt{|\gc Q E'|}}\, .
\end{equation}

Naturally, gauge-field modes whose wavelength is much smaller than the Schwinger scale, $\delta x_{\mathrm{S}}$, will not notice the presence of charged particles and, thus, the corresponding mode functions will not be damped due to the conductivity. Furthermore, one should consider this scale also in the context of the Hubble scale. If $\delta x_{\mathrm{S}} > H^{-1}$, the real fermion--antifermion pairs are seemingly created at separations larger than the Hubble radius at the given moment of time, which collides with our understanding of causality. This is yet another scale-dependent effect that one should consider for realistic modeling of Schwinger damping, i.e., that while $\delta x_{\mathrm{S}} > H^{-1}$, one should consider that no damping should take place at all. This is also in line with the basic assumptions in the derivation of the induced current in Eq.~\eqref{eq: current-colinear-frame}, which is only valid when $|\gc Q E'| \gg H^{2}$.

This motivates a second choice for the scale-dependence function $\Theta(t,k)$. For simplicity, we will model this $k$-dependence of the conductivities by two simple step functions,
\begin{equation}
    \label{eq: beta-k-dep}
    \Theta(t,k) = \theta\big(k_{\mathrm{S}}(t)-k\big) \theta\big(k_{\mathrm{S}}(t)- a(t)H(t)\big) \, ,
\end{equation}
where the threshold momentum $k_{\mathrm{S}}(t)$ corresponds to the Schwinger pair-creation scale in Eq.~\eqref{eq: Schwinger-timescale}. For definiteness, we \textit{define} $k_{\mathrm{S}}(t)$ as
\begin{equation}
    \label{eq: k-S}
    k_{\mathrm{S}}(t)\equiv a(t) \sqrt{|\gc Q\bm{E}'(t)|}=a(t)\frac{|\gc Q|}{2^{1/4}} \left[\bm{E}^2 -\bm{B}^2 + \sqrt{\left(\bm{E}^2 - \bm{B}^2\right)^2 + 4\left(\bm{E}\cdot\bm{B}\right)^2}\right]^{1/4}\, ,
\end{equation}
where we used Eq.~\eqref{eq: Collinear E-prime} for the electric field in the collinear frame.

Therefore, one can separate the time evolution of the gauge-field mode function into two stages. At the first stage, when $k > k_{\mathrm{S}}(t)$, it is described by Eq.~\eqref{eq: Mode Equation} with both electric and magnetic conductivities set to zero. In conformal time, it takes the form
\begin{equation}
    \label{eq: mode-eq without sigma}
    A''_{\lambda}(\eta, k)+(k^2 - 2\lambda \xi k a H)A_{\lambda}(\eta, k)=0.
\end{equation}
Deep inside the horizon, when $k\gg 2aH |\xi|$, the solution must satisfy the Bunch--Davies boundary condition
\begin{equation}
    \label{eq: BDvacuum}
    A_{\lambda}(\eta, k)=\frac{1}{\sqrt{2k}} e^{-ik\eta}\, .
\end{equation}
If we again assume that inflationary expansion of the universe is described by the de Sitter solution $a(\eta)=-1/(H\eta)$ and that the parameter $\xi$ changes adiabatically slowly, then the solution of Eq.~\eqref{eq: mode-eq without sigma} satisfying the Bunch--Davies boundary condition in Eq.~\eqref{eq: BDvacuum} can be expressed in terms of the Whittaker $W$ function:
\begin{equation}
    \label{eq: solution-no-sigma}
    A_{\lambda}(\eta, k)=\frac{1}{\sqrt{2k}} e^{\frac{\lambda \pi \xi}{2}} W_{-i\lambda\xi,\frac{1}{2}}(2ik\eta)\, .
\end{equation}

At the second stage, when $k$ becomes smaller than $k_{\mathrm{S}}(t)$, the mode equation coincides with Eq.~\eqref{eq: Mode Equation}. This is apparent in the damping parameter $\Delta$, which takes the form
\begin{equation}
\label{eq: Delta-new}
    \Delta(t, k)=\exp\Bigg(-\int\limits_{t^{\ast}(k)}^{t} \sigma_E(t') \D t' \Bigg)\, ,
\end{equation}
where $t^{\ast}(k)$ is the moment of time when the mode with wavenumber $k$ starts feeling the conductivity, i.e., the solution of the equation $k=k_{\mathrm{S}}(t)$. Thus, for a given mode, $t^{\ast}(k)$ marks the moment in time that separates the two stages of the evolution.

For a given mode $A_\lambda$ with wavenumber $k$, the solution at $t < t^{\ast}(k)$ is approximately given by Eq.~\eqref{eq: solution-no-sigma}. For times $t > t^{\ast}(k)$, it is instead given by the general solution in Eq.~\eqref{eq: general A mode solution}. The constants $C_{\lambda}$ and $D_{\lambda}$ can be determined by matching the two solutions in Eqs.~\eqref{eq: solution-no-sigma} and \eqref{eq: general A mode solution} at $t = t^{\ast}(k)$. Since the mode function and its time derivative determine the power spectra of the magnetic and electric energy densities, which are the physical observables, the former quantities must be continuous functions of time. Continuity implies the following conditions to be satisfied at the matching point $t=t^{\ast}(k)$:
\begin{subequations}
\label{eq: system-C-D}
    \begin{align}
       W &= C_\lambda \tilde{W} + D_\lambda \tilde{M}\, ,\\
       W'&= C_\lambda \left[\tilde{W}'+ i \frac{a^{\ast} H^{\ast}}{2k}s_E \tilde{W}\right] + D_\lambda \left[\tilde{M}'+ i \frac{a^{\ast} H^{\ast}}{2k}s_E \tilde{M}\right]\, ,
    \end{align}
\end{subequations}
where 
\begin{subequations}
\begin{align}
    W & = W_{-i\lambda\xi,\frac{1}{2}}\big(-\frac{2ik}{a^\ast H^\ast}\big) \,, \\
    \tilde{W} & = W_{-i\lambda\xi_{\mathrm{eff}},\frac{1}{2}+s_E}\big(-\frac{2ik}{a^\ast H^\ast}\big) \,, \\
    \tilde{M} & = M_{-i\lambda\xi_{\mathrm{eff}},\frac{1}{2}+s_E}\big(-\frac{2ik}{a^\ast H^\ast}\big) \,,
\end{align}%
\end{subequations}
and where a prime denotes the derivative of a Whittaker function with respect to its argument, i.e., $W'_{\kappa,\mu}(z)=\frac{d}{dz}W_{\kappa,\mu}(z)$. Symbols with an asterisk denote the corresponding quantities evaluated at the moment $t^\ast (k)$.
Solving the system of linear algebraic equations~\eqref{eq: system-C-D}, we determine the coefficients $C_\lambda(k)$ and $D_\lambda(k)$:
\begin{align}
\label{eq: C-lambda}
    C_\lambda(k)&=i\frac{\Gamma(1+s_E + i\lambda\xi_{\mathrm{eff}})}{\Gamma(2+2s_E)}\frac{a^\ast \!H^\ast}{2k}\!\left[W_{+1}\tilde{M}\!+\!(1+s_E - i\lambda\xi_{\mathrm{eff}}) W \tilde{M}_{+1}+(s_E + i\lambda s_B)W \tilde{M}\right]\, ,\\
\label{eq: D-lambda}
    D_\lambda(k)&=-i\frac{\Gamma(1+s_E + i\lambda\xi_{\mathrm{eff}})}{\Gamma(2+2s_E)}\frac{a^\ast\! H^\ast}{2k}\!\left[W_{+1}\tilde{W}- W \tilde{W}_{+1}+(s_E + i\lambda s_B)W \tilde{W}\right]\, ,
\end{align}
where the subscript $+1$ denotes the corresponding Whittaker function with the index $\kappa$ increased by one unit, e.g., $W_{+1}=W_{1-i\lambda\xi,\frac{1}{2}}\big(-\frac{2ik}{a^\ast H^\ast}\big)$, which follows from the derivatives of the Whittaker functions. The Euler $\Gamma$ functions in these expressions originate from the Wronskian of the Whittaker functions $W$ and $M$ which turns out to be the determinant of the coefficient matrix of the linear system in Eq.~\eqref{eq: system-C-D}. The properties of the Whittaker function that we used in the derivation of Eqs.~\eqref{eq: C-lambda} and \eqref{eq: D-lambda} are listed in appendix~\ref{app: Whittaker}.


\section{Gradient-expansion formalism}
\label{sec:GEF}


\subsection{GEF with scale-dependent Schwinger damping}
\label{subsec:GEF-scale-dep}

Having discussed our improved modeling of Schwinger conductivity during axion inflation, we now turn to our method to solve the dynamics of the system given in Eqs.~\eqref{eq: EoMs}. As demonstrated in Ref.~\cite{Gorbar:2021rlt}, this is elegantly done using the gradient-expansion formalism (GEF), in which the equations of motion, Eqs.~\eqref{eq: EoMs}, are re-expressed in terms of the quantum expectation values for bilinear combinations of electric- and magnetic-field operators,
\begin{subequations}
    \begin{align}
        \curly{E}{n} &= \frac{1}{a^n}\braket{\bm{E} \cdot \rot^n \bm{E}}\, , \\
        \curly{G}{n} &= -\frac{1}{2a^n}[\braket{\bm{B} \cdot \rot^n \bm{E}} + \braket{\bm{E} \cdot \rot^n \bm{B}}]\, ,\\
        \curly{B}{n} &= \frac{1}{a^n}\braket{\bm{B} \cdot \rot^n \bm{B}}\, .
    \end{align}
    \label{eq: Bilinear Terms}%
\end{subequations}%
These expectation values are readily computed by expressing the electric and magnetic fields in terms of the four-vector potential, Eq.~\eqref{eq: EandBthroughF}, and decomposing the gauge-field operator in terms of mode functions $A_\lambda(t,k)$ with definite wavenumber $k$ and circular polarization $\lambda=\pm$, as in Eq.~\eqref{eq: FourierModes}:
\begin{subequations}
    \begin{align}
        \curly{E}{n}(t) &=  \int\limits_0^{k_{\mathrm{h}}(t)} \frac{\D  k}{k}  \frac{k^{n+3}}{2 \pi^{2}a^{n+2}} \sum_{\lambda}\lambda^n |\dot{A}_\lambda(t,k)|^2\, , \\
        \curly{G}{n}(t) &=  \int\limits_0^{k_{\mathrm{h}}(t)} \frac{\D  k}{k}  \frac{k^{n+4}}{2 \pi^{2}a^{n+3}} \sum_{\lambda}\lambda^{n+1} \mathrm{Re}[\dot{A}_\lambda(t,k)A_\lambda^*(t,k)]\, ,\\
        \curly{B}{n}(t) &= \int\limits_0^{k_{\mathrm{h}}(t)} \frac{\D  k}{k}  \frac{k^{n+5}}{2 \pi^{2}a^{n+4}} \sum_{\lambda}\lambda^n |A_\lambda(t,k)|^2\, . 
    \end{align}
    \label{eq: Bilinear Terms Integrals}%
\end{subequations}%
Here, we regularized the UV-divergent integrals by means of an explicit cutoff scale, $k_{\mathrm{h}}(t)$. As the UV divergence originates from an infinite integral over vacuum modes of the gauge field, a natural cutoff scale for the integral is the momentum $k_{\mathrm{crit}}$ given in Eq.~\eqref{eq: instability scale}. At a given moment in time, all modes with wavenumber $k \leq k_{\mathrm{crit}}$ have experienced the tachyonic instability and are correspondingly no longer in a vacuum configuration. From Eq.~\eqref{eq: instability scale}, we can therefore determine that the cutoff scale $k_{\mathrm{h}}(t)$ can naturally be chosen as
\begin{equation}
    k_{\mathrm{h}}(t) = \max_{t' \leq t} \bigg\{ a(t') H(t') \bigg[ |\xi_{\mathrm{eff}}(t')| + \sqrt{\xi_{\mathrm{eff}}^2(t') + s_E^2(t') + s_E(t')} \bigg] \bigg\}\, .
    \label{eq: cut-off scale}
\end{equation}
Here, we allow for a general time dependence of the scale factor $a$ and the Hubble rate $H$. Furthermore, the parameters $\xi_{\mathrm{eff}}$ and $s_E$, which we introduced in Eq.~\eqref{eq: gauge-field production parameters}, are themselves generally time-dependent. Therefore, the cutoff scale $k_{\mathrm{h}}(t)$ is itself a time-dependent quantity. Physically, this time dependence is due to the expansion of the Universe, which implies that over time more gauge modes are excited out of their vacuum state as they start experiencing the presence of the axion field. The maximum over all preceding times is taken in order to ensure that $k_{\mathrm{h}}(t)$ is a monotonically increasing function.

We are now in a position to construct the dynamical equations for the quantum expectation values in Eq.~\eqref{eq: Bilinear Terms} directly from the modified mode equation \eqref{eq: Mode Equation, K-dep Sigmas}. For now, we keep the scale-dependence function $\Theta(t,k)$ arbitrary. In a second step, we will discuss the two scenarios defined through $\Theta(t,k) = 1$ and $\Theta(t,k) = \theta\big(k_{\mathrm{S}}(t)-k\big)\theta\big(k_{\mathrm{S}}(t)- a(t)H(t)\big)$ as well as a simplifying approximation of the latter. The GEF equations in this general setup are
\begin{equation}
        \ddot{\phi} + 3H \dot{\phi} + V_{,\phi} = - I_{,\phi} \curly{G}{0}\, , \label{eq: GEF general - KG}
    \end{equation}
\begin{equation}
        H^2 = \frac{1}{3 M_{\mathrm{P}}^2} \left[\frac{1}{2}\dot{\phi}^2 + V(\phi) + \frac{1}{2} \curly{E}{0} + \frac{1}{2} \curly{B}{0} + \rho_\chi\right]\, , \label{eq: GEF General - Friedmann}
    \end{equation}
\begin{subequations}
    \begin{equation}
        \curly{\dot{E}}{n} + (4+n)H \curly{E}{n}  + 2\curly{G}{n+1} - 2 I_{,\phi}\dot{\phi} \curly{G}{n} + 2\sigma_E \barcurly{E}{n}- 2 \sigma_B \barcurly{G}{n} =  S_{\mathcal{E}}^{(n)}\, ,\label{eq: GEF General - En}
    \end{equation}
    \begin{equation}
        \curly{\dot{G}}{n} + (4+n)H \curly{G}{n} - \curly{E}{n+1} + \curly{B}{n+1} - I_{,\phi}\dot{\phi} \curly{B}{n} + \sigma_E \barcurly{G}{n} - \sigma_B \barcurly{B}{n} = S_{\mathcal{G}}^{(n)}\, , \label{eq: GEF General - Gn}
    \end{equation}
    \begin{equation}
        \curly{\dot{B}}{n} + (4+n) H \curly{B}{n} - 2\curly{G}{n+1}  =  S_{\mathcal{B}}^{(n)}\, , \label{eq: GEF General - Bn}
    \end{equation}
    \label{eq: GEF General}%
\end{subequations}
\begin{equation}
        \dot{\rho}_\chi + 4 H \rho_\chi = (\sigma_E \barcurly{E}{0} - \sigma_B \barcurly{G}{0}) \, . \label{eq: GEF General - rho-chi}
\end{equation}
The last equation is included in order to obtain a closed system of equations, which requires knowledge on the evolution of the fermion energy density, $\rho_\chi$. The equation was derived by considering global energy conservation in the expanding universe. The details of this derivation can be found in Appending \ref{app: rhoChi EoM}.

The source terms on the right-hand side of Eq.~\eqref{eq: GEF General} appear due to the time-dependent UV cutoff in Eq.~\eqref{eq: cut-off scale}. They are given by
\begin{subequations}
\begin{align}
    \label{eq: boundary-En}
    S_{\mathcal{E}}^{(n)}&=\frac{d \ln k_{\mathrm{h}}(t)}{d t}\frac{\Delta\big(t, k_{\mathrm{h}}(t)\big)}{4\pi^{2}}\left(\frac{k_{\mathrm{h}}(t)}{a(t)}\right)^{n+2}\sum_{\lambda=\pm 1}\lambda^{n} 2k_{\mathrm{h}}(t)\big|\dot{A}_\lambda\big(t,k_{\mathrm{h}}(t)\big)\big|^2\, ,\\
    \label{eq: boundary-Gn}
    S_{\mathcal{G}}^{(n)}&=\frac{d \ln k_{\mathrm{h}}(t)}{d t}\frac{\Delta\big(t, k_{\mathrm{h}}(t)\big)}{4\pi^{2}}\left(\frac{k_{\mathrm{h}}(t)}{a(t)}\right)^{n+3}\sum_{\lambda=\pm 1}\lambda^{n+1} 2k_{\mathrm{h}}(t) \operatorname{Re}\big[A^{\ast}_\lambda\big(t,k_{\mathrm{h}}(t)\big)\dot{A}_\lambda\big(t,k_{\mathrm{h}}(t)\big)\big]\, ,\\
    \label{eq: boundary-Bn}
    S_{\mathcal{B}}^{(n)}&=\frac{d \ln k_{\mathrm{h}}(t)}{d t}\frac{\Delta\big(t, k_{\mathrm{h}}(t)\big)}{4\pi^{2}}\left(\frac{k_{\mathrm{h}}(t)}{a(t)}\right)^{n+4}\sum_{\lambda=\pm 1}\lambda^{n} 2k_{\mathrm{h}}(t) \big|A_\lambda\big(t,k_{\mathrm{h}}(t)\big)\big|^2\, ,
\end{align}
\label{eq: Source Terms}%
\end{subequations}
where $A_\lambda\big(t,k_{\mathrm{h}}(t)\big)$ is the gauge-field mode function whose momentum equals the threshold momentum $k_{\mathrm{h}}(t)$ at this moment of time, i.e., $k=k_{\mathrm{h}}(t)$.

Note that our GEF equations contain novel terms compared to previous results in the literature, which are the quantities $\barcurly{X}{n}$, where $\mathcal{X} = \mathcal{E}$, $ \mathcal{G}$ or $\mathcal{B}$. They appear due to the scale-dependent prefactors $\Theta(t,k)$ that quantify the scale dependence of the conductivities $\sigma_{E/B}$. They are expressed in terms of the mode functions $A_\lambda$ through
\begin{subequations}
\begin{align}
    \label{eq: P-En}
    \barcurly{E}{n}&=  \int\limits_0^{k_{\mathrm{h}}(t)} \frac{\D  k}{k}  \frac{k^{n+3}}{2 \pi^{2}a^{n+2}}\, \Theta(t,k) \sum_{\lambda}\lambda^n |\dot{A}_\lambda(t,k)|^2\, , \\
    \label{eq: P-Gn}
    \barcurly{G}{n}&= \int\limits_0^{k_{\mathrm{h}}(t)} \frac{\D  k}{k}  \frac{k^{n+4}}{2 \pi^{2}a^{n+3}}\, \Theta(t,k) \sum_{\lambda}\lambda^{n+1} \mathrm{Re}[\dot{A}_\lambda(t,k)A_\lambda^*(t,k)]\, , \\
    \label{eq: P-Bn}
    \barcurly{B}{n}&= \int\limits_0^{k_{\mathrm{h}}(t)} \frac{\D  k}{k}  \frac{k^{n+5}}{2 \pi^{2}a^{n+4}}\, \Theta(t,k) \sum_{\lambda}\lambda^n |A_\lambda(t,k)|^2\, .
 \end{align}
\label{eq: Partial Damping Terms}%
\end{subequations}
The interpretation of these terms is evident: they account for the fact that only some modes contributing to the expectation values are damped by the conductive medium.

To close the system of equations, let us now return to the results of section~\ref{subsec: Sigmas} and express them in terms of the GEF quantities $\curly{E}{n}$, $\curly{G}{n}$, and $\curly{B}{n}$. Taking into account the quantum nature of the gauge fields and replacing all scalar products of $\bm{E}$ and $\bm{B}$ in Eqs.~\eqref{eq: sigma-E-B} by their quantum expectation values, we obtain the electric and magnetic conductivities in the form
\begin{subequations}
    \begin{align}
        \sigma_E &= \frac{|\bm{J}'|}{\sqrt{\Sigma}} \sqrt{\frac{\curly{E}{0} - \curly{B}{0} + \Sigma}{\curly{E}{0} + \curly{B}{0} + \Sigma}}\, ,\\
        \sigma_B &=  - \operatorname{sign} \, \curly{G}{0} \frac{|\bm{J}'|}{\sqrt{\Sigma}} \sqrt{\frac{\curly{B}{0} - \curly{E}{0} + \Sigma}{\curly{E}{0} + \curly{B}{0} + \Sigma}}\, ,
    \end{align}  
\end{subequations}
where we defined
\begin{equation}
    \Sigma = \sqrt{\left(\curly{E}{0} - \curly{B}{0}\right)^2 + 4 \left(\curly{G}{0}\right)^2}\, .
\end{equation}
Using the explicit expression for the Schwinger-induced current, Eq.~\eqref{eq: current in collinear frame 2}, in the collinear frame and treating all fermions as massless particles gives
\begin{equation}
    |\bm{J}'| = C \frac{\gc^3}{6 \pi^2 H} E' |B'| \coth{\left(\pi \frac{|B'|}{|E'|}\right)},
\end{equation}
where the prefactor $C$ takes into account contributions from different fermion species in the plasma. Expressing everything in terms of $\curly{E}{n}$, $\curly{B}{n}$, and $\curly{G}{n}$, we finally arrive at the effective conductivities in the mixed picture:
\begin{subequations}
    \begin{align}
        \sigma_E &= C \frac{\gc^3}{6 \pi^2 H}\frac{|\curly{G}{0}|}{\sqrt{\Sigma}} \sqrt{\frac{\curly{E}{0} - \curly{B}{0} + \Sigma}{\curly{E}{0} + \curly{B}{0} + \Sigma}} \,\coth{\left(\pi \sqrt{\frac{\curly{B}{0} - \curly{E}{0} + \Sigma}{\curly{E}{0} - \curly{B}{0} + \Sigma}}\right)}\, ,\\
        \sigma_B &= -C \frac{\gc^3}{6 \pi^2 H}\frac{\curly{G}{0}}{\sqrt{\Sigma}} \sqrt{\frac{\curly{B}{0} - \curly{E}{0} + \Sigma}{\curly{E}{0} + \curly{B}{0} + \Sigma}} \,\coth{\left(\pi \sqrt{\frac{\curly{B}{0} - \curly{E}{0} + \Sigma}{\curly{E}{0} - \curly{B}{0} + \Sigma}}\right)}\, .
    \end{align}
\label{eq: sigma-E-B-gef-mixed}%
\end{subequations}
These expressions represent some of the main new results that we present in this paper.

For completeness, let us also show the expressions for the Schwinger conductivities in the pictures that were used in the literature before. For the electric picture, we have
\begin{equation}
\label{eq: sigma-E-B-gef-electric}
    \sigma_{E} = C \frac{\gc^3}{6 \pi^2 H} \sqrt{\curly{B}{0}} \coth{\left(\pi \sqrt{\frac{\curly{B}{0}}{\curly{E}{0}}}\right)}\, ,\qquad \sigma_B=0\, ,
\end{equation}
while for the magnetic picture
\begin{equation}
\label{eq: sigma-E-B-gef-magnetic}
    \sigma_E=0\, , \qquad \sigma_{B} = -C \operatorname{sign}\big(\curly{G}{0}\big) \frac{\gc^3}{6 \pi^2 H} \sqrt{\curly{E}{0}} \coth{\left(\pi \sqrt{\frac{\curly{B}{0}}{\curly{E}{0}}}\right)}\, .
\end{equation}
We compare the numerical results obtained in these three pictures in section~\ref{sec:results}.

\subsection{A tale of three scale dependencies}
\label{subsec:different-scale-dependencies}

Having derived the general GEF equations for arbitrary $\Theta(t,k)$, let us now discuss the three different choices for this function that we will use in our numerical analysis.

\subsubsection{No scale dependence}
\label{subsubsec: no scale dependence}

First, consider the case of Schwinger damping on all scales, i.e., $\Theta(t,k) = 1$. As evident from Eqs.~\eqref{eq: Bilinear Terms Integrals} and \eqref{eq: Partial Damping Terms}, this case immediately implies that $\barcurly{X}{n} = \curly{X}{n}$ for $\mathcal{X} = \mathcal{E}$, $\mathcal{G}$, or $\mathcal{B}$. Furthermore, to construct the source terms $S_\mathcal{X}^{(n)}$ in Eqs.~\eqref{eq: Source Terms}, we use that the mode function $A_\lambda$ for the mode with wavenumber $k=k_{\mathrm{h}}(t)$ is approximately given by Eq.~\eqref{eq: solution-no-scale-dep}. This is justified as we only need to know the mode function when the corresponding mode enters the tachyonic-instability region. Therefore, we can take the values of parameters $\xi_{\mathrm{eff}}$ and $s_E$ at this moment of time and assume that they are constant in the vicinity of this moment. 
The GEF equations, Eqs.~\eqref{eq: GEF General} and \eqref{eq: Source Terms}, in this scenario are identical to the ones of Ref.~\cite{Gorbar:2021rlt} if we work in the electric picture, $\sigma_B = 0$ (see Eq.~\eqref{eq: sigma-E-B-gef-electric}). In the magnetic and mixed picture, an additional magnetic conductivity term $\sigma_B$ appears, which is not present in Ref.~\cite{Gorbar:2021rlt}. This magnetic-conductivity term can be obtained from the equations in Ref.~\cite{Gorbar:2021rlt} via the replacement rule $I_{,\phi}\dot\phi \to (I_{,\phi}\dot\phi + \sigma_B)$ inhibiting the coupling of the gauge field to the inflaton (note that the signs of $\sigma_B$ and $I_{,\phi}\dot\phi$ are typically opposite to each other).

\subsubsection[Damping on scales \texorpdfstring{$k\leq k_{\mathrm{S}}(t)$}{k<=kS(t)}]{Damping on scales \texorpdfstring{\boldmath{$k\leq k_{\mathrm{S}}(t)$}}{k<=kS(t)}}
\label{subsubsec: true scale dep}

\begin{figure}
    \centering
    \includegraphics[width=0.49\textwidth]{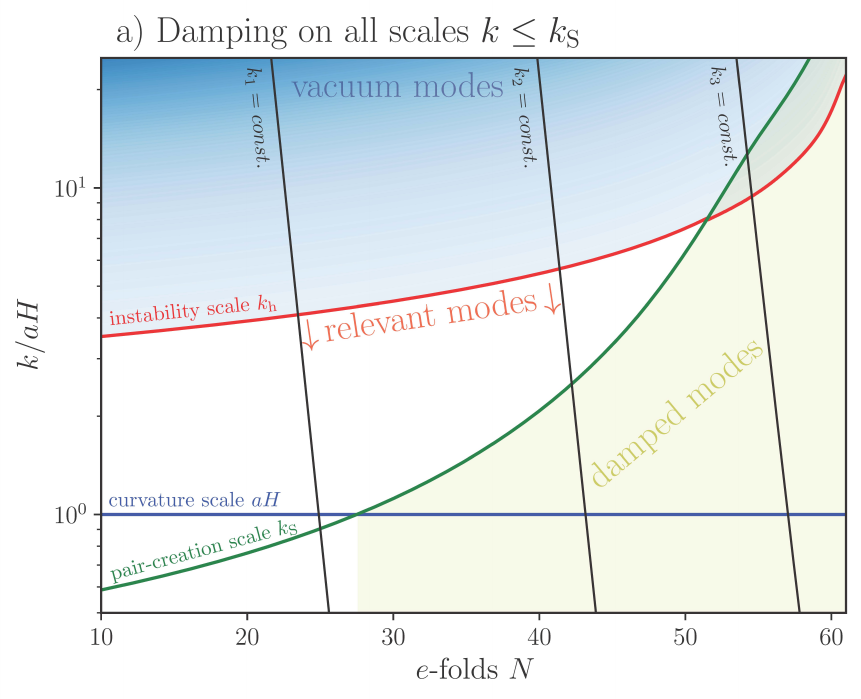}
    \hfill
    \includegraphics[width=0.49\textwidth]{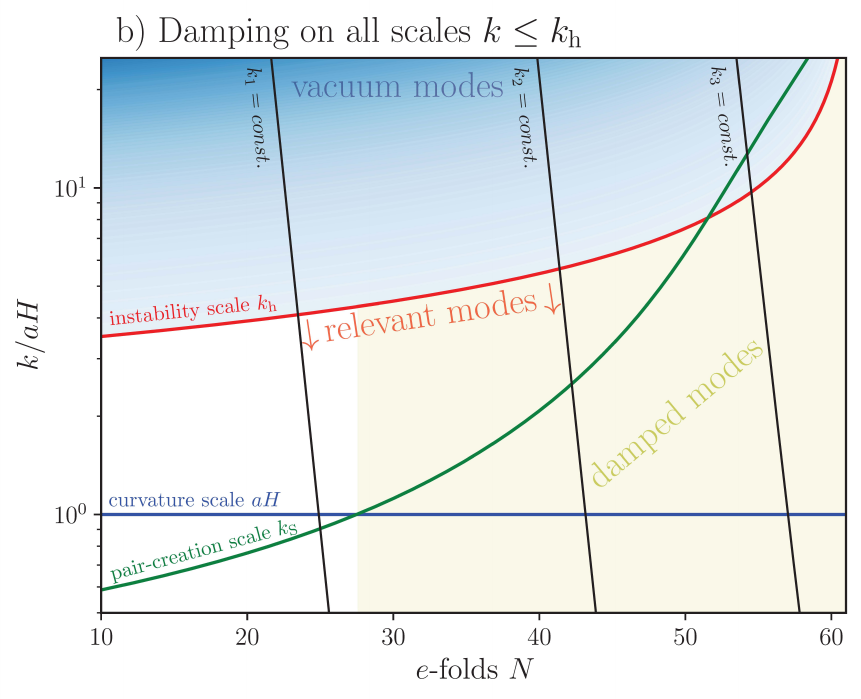}
    \caption{Time evolution of the three relevant scales during inflation: Hubble scale $aH$ (blue), Schwinger pair-creation scale $k_{\mathrm{S}}$ (green), and tachyonic-instability scale $k_{\mathrm{h}}$ (red). The blue shaded region shows the gauge-field modes that have not yet experienced the tachyonic instability and which are thus still close to the vacuum state. The green shaded region denotes the modes that undergo Schwinger damping. \textbf{Panel~(a)} corresponds to the real physical case when the damping occurs for all modes with momenta $k\leq k_{\mathrm{S}}(t)$ once $a(t)H(t)<k_{\mathrm{S}}(t)$. Three tilted black lines correspond to gauge-field modes with different scenarios of evolution: the leftmost mode with momentum $k_1$ is not affected by the conductivity until the Schwinger effect sets in (until it crosses the left vertical boundary of the green shaded region); the mode in the middle ($k_2$) first experiences the tachyonic instability and then Schwinger damping; the rightmost mode ($k_3$) undergoes damping prior to the tachyonic enhancement. \textbf{Panel~(b)} represents the approximation when the damping is turned on for modes with $k\leq k_{\mathrm{h}}(t)$ once $a(t)H(t)<k_{\mathrm{S}}(t)$. Here, the mode $k_1$ evolves in the same way as the one in panel~(a), while the modes with momenta $k_2$ and $k_3$ undergo damping when they become tachyonically unstable. \label{fig: illustration-scale-dependence}}
\end{figure}

Next, consider our novel approach to model the scale dependence through
\begin{equation}
    \Theta(t,k) = \theta\big(k_{\mathrm{S}}(t)-k\big) \theta\big(k_{\mathrm{S}}(t)- a(t)H(t)\big)\, .
\end{equation}
By construction of this function, we can see that there are three relevant scales in this scenario: the comoving Hubble scale $a(t)H(t)$, the tachyonic-instability scale $k_{\mathrm{h}}(t)$ and the Schwinger-damping scale, $k_{\mathrm{S}}(t)$. The latter is given by Eq.~\eqref{eq: k-S}, which, expressed in terms of GEF bilinear quantities, reads
\begin{equation}
\label{eq: k-S-GEF}
    k_{\mathrm{S}}(t)=a\frac{C^{1/3}|\gc|}{2^{1/4}} \left[\curly{E}{0} -\curly{B}{0} + \sqrt{\left(\curly{E}{0} -\curly{B}{0}\right)^2 + 4\left(\curly{G}{0}\right)^2}\right]^{1/4}\, .
\end{equation}
The evolution of a mode $A_\lambda(t,k)$ for a given wavenumber $k$ is affected by the ordering of these three scales at a given time $t$; see figure~\ref{fig: illustration-scale-dependence}(a). As long as $k_{\mathrm{S}}(t) < a(t)H(t)$, no mode feels the presence of the conductive medium, as we consider that the electric field is not yet strong enough to effectively produce real fermion--antifermion pairs. This is illustrated by the mode labeled $k_1$ in figure~\ref{fig: illustration-scale-dependence}. Correspondingly, no damped modes contribute to the expectation values $\curly{X}{n}$ for $\mathcal{X}=\mathcal{E}$, $\mathcal{G}$ or $\mathcal{B}$ in Eq.~\eqref{eq: Bilinear Terms Integrals}. Once $a(t)H(t) < k_{\mathrm{S}}(t)$, a mode of wavenumber $k$ will experience damping as soon as $k \leq k_{\mathrm{S}}(t)$. But as long as $k_{\mathrm{S}}(t) < k_{\mathrm{h}}(t)$, any mode will first experience the tachyonic instability before becoming damped. So only some modes contributing to $\curly{X}{n}$ will feel the damping medium. This is illustrated by the mode labeled $k_2$ in figure~\ref{fig: illustration-scale-dependence}(a). Only when $k_{\mathrm{h}}(t) \leq k_{\mathrm{S}}(t)$ will a mode experience damping before feeling the tachyonic instability. In this case, all modes entering $\curly{X}{n}$ are Schwinger-damped. This is illustrated by the mode labeled $k_3$ in figure~\ref{fig: illustration-scale-dependence}(a).

After these qualitative observations, let us now study how these effects are captured by the GEF equations, Eqs.~\eqref{eq: GEF General}. The Heaviside theta functions in Eqs.~\eqref{eq: Partial Damping Terms} imply that
\begin{subequations}
 \begin{align}
    \barcurly{E}{n}&=\!\!\!\!  \int\limits_{0}^{\min(k_{\mathrm{S}}, k_{\mathrm{h}})}\!\!\!\! \frac{\D  k}{k}  \frac{k^{n+3}}{2 \pi^{2}a^{n+2}} \sum_{\lambda}\lambda^n |\dot{A}_\lambda(t,k)|^2\theta_{\mathrm{S}} \equiv \theta_{\mathrm{S}}\,\tildecurly{E}{n}\, ,\\
    \barcurly{G}{n}&=\!\!\!\! \int\limits_0^{\min(k_{\mathrm{S}}, k_{\mathrm{h}})}\!\!\!\! \frac{\D  k}{k}  \frac{k^{n+4}}{2 \pi^{2}a^{n+3}} \sum_{\lambda}\lambda^{n+1} \mathrm{Re}[\dot{A}_\lambda(t,k)A_\lambda^*(t,k)]\theta_{\mathrm{S}}\equiv \theta_{\mathrm{S}}\,\tildecurly{G}{n}\, , \\
    \barcurly{B}{n}&= \!\!\!\!\int\limits_{0}^{\min(k_{\mathrm{S}}, k_{\mathrm{h}})} \!\!\!\! \frac{\D  k}{k}  \frac{k^{n+5}}{2 \pi^{2}a^{n+4}} \sum_{\lambda}\lambda^n |A_\lambda(t,k)|^2 \theta_{\mathrm{S}} \equiv \theta_{\mathrm{S}}\,\tildecurly{B}{n} \, ,
 \end{align}
\label{eq: Partial Damping Terms scale Dep}%
\end{subequations}
where $\theta_{\mathrm{S}} = \theta(k_{\mathrm{S}}- aH)$ and where we suppressed the time dependence of $k_{\mathrm{S}}$, $k_{\mathrm{h}}$, $a$, and $H$.

We notice that the expressions for $\tildecurly{X}{n}$ are completely analogous to the integral expressions for $\curly{X}{n}$ in Eq.~\eqref{eq: Bilinear Terms Integrals}, except that, while $k_{\mathrm{S}} < k_{\mathrm{h}}$, the cutoff momentum for the integrals is given by $k_{\mathrm{S}}$ not $k_{\mathrm{h}}$ since only modes below $k_{\mathrm{S}}$ can experience the damping by the conductive medium. When $k_{\mathrm{h}} \leq k_{\mathrm{S}}$, all modes entering $\curly{X}{n}$ have already experienced the damping medium; correspondingly $\barcurly{X}{n} = \curly{X}{n}$.
Furthermore, the factor $\theta_{\mathrm{S}} = \theta(k_{\mathrm{S}}- aH)$ ensures that damping is disregarded for weak electric fields whose Schwinger-damping length scale is estimated to be larger than the Hubble radius. These results can be clearly understood in terms of the qualitative picture presented in figure~\ref{fig: illustration-scale-dependence}(a).

To compute the time evolution of $\tildecurly{X}{n}$ while $k_{\mathrm{S}}(t) < k_{\mathrm{h}}(t)$, we can formulate a completely new set of GEF equations for the quantities $\tildecurly{X}{n}$ that need to be solved together with the original set of differential equations in Eqs.~\eqref{eq: GEF general - KG} to \eqref{eq: GEF General - rho-chi}. We can construct the differential equations for $\tildecurly{X}{n}$ by again using the mode equation \eqref{eq: Mode Equation, K-dep Sigmas}. The derivation is completely analogous to the one used to derive Eqs.~\eqref{eq: GEF General}. They read
\begin{subequations}
\begin{equation}
        \curly{\dot{\tilde{E}}}{n} + \left[(4+n)H + 2\sigma_E \right] \tildecurly{E}{n} + 2\tildecurly{G}{n+1} - 2 I_{,\phi}\dot{\phi} \tildecurly{G}{n} - 2 \sigma_B \tildecurly{G}{n} =  \tilde{S}_{\mathcal{E}}^{(n)}\, ,\label{eq: GEF - TildeEn}
\end{equation}
\begin{equation}
        \curly{\dot{\tilde{G}}}{n} + \left[(4+n)H + \sigma_E\right] \tildecurly{G}{n} - \tildecurly{E}{n+1} + \tildecurly{B}{n+1} - I_{,\phi}\dot{\phi} \tildecurly{B}{n} - \sigma_B \tildecurly{B}{n} = \tilde{S}_{\mathcal{G}}^{(n)}\, , \label{eq: GEF - TildeGn}
\end{equation}
\begin{equation}
        \curly{\dot{\tilde{B}}}{n} + (4+n) H \tildecurly{B}{n} - 2\tildecurly{G}{n+1}  =  \tilde{S}_{\mathcal{B}}^{(n)}\, , \label{eq: GEF - TildeBn}
\end{equation}%
\label{eq: GEF - Tilde}%
\end{subequations}%
with the source terms given by
\begin{subequations}
\begin{align}
    \label{eq: boundary-TildeEn}
    \tilde{S}_{\mathcal{E}}^{(n)}&=\frac{d \ln k_{\mathrm{S}}(t)}{d t}\frac{1}{4\pi^{2}}\left(\frac{k_{\mathrm{S}}(t)}{a(t)}\right)^{n+2}\sum_{\lambda=\pm 1}\lambda^{n} 2k_{\mathrm{S}}(t)\big|\dot{A}_\lambda\big(t,k_{\mathrm{S}}(t)\big)\big|^2\, ,\\
    \label{eq: boundary-TildeGn}
    \tilde{S}_{\mathcal{G}}^{(n)}&=\frac{d \ln k_{\mathrm{S}}(t)}{d t}\frac{1}{4\pi^{2}}\left(\frac{k_{\mathrm{S}}(t)}{a(t)}\right)^{n+3}\sum_{\lambda=\pm 1}\lambda^{n+1} 2k_{\mathrm{S}}(t) \operatorname{Re}\big[A^{\ast}_\lambda\big(t,k_{\mathrm{S}}(t)\big)\dot{A}_\lambda\big(t,k_{\mathrm{S}}(t)\big)\big]\, ,\\
    \label{eq: boundary-TildeBn}
    \tilde{S}_{\mathcal{B}}^{(n)}&=\frac{d \ln k_{\mathrm{S}}(t)}{d t}\frac{1}{4\pi^{2}}\left(\frac{k_{\mathrm{S}}(t)}{a(t)}\right)^{n+4}\sum_{\lambda=\pm 1}\lambda^{n} 2k_{\mathrm{S}}(t) \big|A_\lambda\big(t,k_{\mathrm{S}}(t)\big)\big|^2\, .
\end{align}%
\label{eq: Source Terms for kS}%
\end{subequations}%
The origin of these source terms is similar to those in Eq.~\eqref{eq: Source Terms}. They arise due to the time dependence of the upper integration limit, $k_{\mathrm{S}}(t)$ in Eq.~\eqref{eq: Partial Damping Terms scale Dep}. Correspondingly, they account for the fact that, over time, more modes experience the damping conductive medium.

Let us turn our attention to the practical evaluation of the source terms $S_{\mathcal{X}}^{(n)}$ and $\tilde{S}_{\mathcal{X}}^{(n)}$ in Eqs.~\eqref{eq: Source Terms} and \eqref{eq: Source Terms for kS}.
The evaluation of the source terms $S_{\mathcal{X}}^{(n)}$ can be understood from figure~\ref{fig: illustration-scale-dependence}(a). While $k_{\mathrm{S}}(t) < a(t)H(t)$, there is no conductive medium yet to be felt by any of the modes. Therefore, all modes, especially the mode $A_{\lambda}\big(t,k_{\mathrm{h}}(t)\big)$ with wavevector $k = k_{\mathrm{h}}(t)$, are approximately given by Eq.~\eqref{eq: solution-no-sigma}. Once fermion production becomes effective, after $a(t)H(t) < k_{\mathrm{S}}(t)$ holds true, the last mode to have experienced the tachyonic instability, namely the one with wavenumber $k=k_{\mathrm{h}}(t)$, still has a wavelength smaller than the typical fermion separation, i.e., $k_{\mathrm{S}}(t) < k_{\mathrm{h}}(t)$. Therefore, $A_{\lambda}\big(t,k_{\mathrm{h}}(t)\big)$ is still given by Eq.~\eqref{eq: solution-no-sigma}. The situation only changes once $k_{\mathrm{h}}(t) \leq k_{\mathrm{S}}(t)$. Now, the evolution of the mode with wavenumber $k=k_{\mathrm{h}}(t)$ is such that it had initially evolved without feeling the presence of the conductive medium, and only later, very suddenly, it starts experiencing the damping. At the point in time when this mode experiences the tachyonic instability, $A_{\lambda}\big(t,k_{\mathrm{h}}(t)\big)$ is given by Eq.~\eqref{eq: general A mode solution} with the coefficients $C_\lambda(k), \, D_\lambda(k)$ given in Eqs.~\eqref{eq: C-lambda} and \eqref{eq: D-lambda} (see the discussion at the end of section~\ref{subsec: Delta}).

To evaluate the second type of source term, $\tilde{S}_{\mathcal{X}}^{(n)}$, we can safely assume that the mode with wavenumber $k_{\mathrm{S}}(t)$ has never experienced damping before, as Eq.~\eqref{eq: Source Terms for kS} is only ever evaluated when $k_{\mathrm{S}}(t) < k_{\mathrm{h}}(t)$. Therefore, we use the approximate analytical solution for $A_{\lambda}\big(t,k_{\mathrm{S}}(t)\big)$ given by equation \eqref{eq: solution-no-sigma}.

\subsubsection[Damping on scales \texorpdfstring{$k\leq k_{\mathrm{h}}(t)$}{k<=kh(t)}]{Damping on scales \texorpdfstring{\boldmath{$k\leq k_{\mathrm{h}}(t)$}}{k<=kh(t)}}
\label{subsubsec: scale dep kh}

The physically motivated model of scale dependence of the Schwinger conductivities described in section~\ref{subsubsec: true scale dep} appears to be rather challenging for numerical implementation because one has to properly track the evolution of all relevant scales and, depending on their interrelations, switch between different definitions of the coefficients $\barcurly{X}{n}$ and boundary terms $S_{\mathcal{X}}^{(n)}$ in the GEF equations. Therefore, we would like to present here an approximate model that incorporates scale dependence while being much simpler in its numerical implementation. As we will see later on, this approximation yields results that are extremely close to the model discussed above; see section~\ref{sec:results}. In this model, the function $\Theta(t,k)$ has the form
\begin{equation}
    \Theta(t,k) = \theta\big(k_{\mathrm{h}}(t)-k\big) \theta\big(k_{\mathrm{S}}(t)- a(t)H(t)\big)\, .
\end{equation}
This choice of $\Theta(t,k)$ evokes that, as soon as the produced gauge field is sufficiently strong for the Schwinger effect to be in operation, $a(t)H(t) < k_{\mathrm{S}}(t)$, every mode starts feeling Schwinger damping at the same time when it undergoes the tachyonic instability. In other words, the vacuum-like modes are not damped while \textit{all} modes outside the instability horizon feel the presence of the conducting medium. This situation is illustrated by figure~\ref{fig: illustration-scale-dependence}(b). Here, the green shaded region is complementary to the blue shaded area once $a(t)H(t) < k_{\mathrm{S}}(t)$ [cf.\ panel~(a) of the same figure for the case of the actual scale dependence].

In this simplified case, we always have
\begin{equation}
   \barcurly{X}{n}(t) = \curly{X}{n}(t) \theta\big(k_{\mathrm{S}}(t)-a(t)H(t)\big)
\end{equation}
for $\mathcal{X}=\mathcal{E},\ \mathcal{G}$, and $\mathcal{B}$ because any mode contributing to the $\curly{X}{n}$ feels the damping medium. Concerning the boundary terms, $S_{\mathcal{X}}^{(n)}$, they are always computed using the mode function $A_\lambda(t,k)$ given by Eq.~\eqref{eq: solution-no-sigma}, which is not damped.


\section{Results and discussion}
\label{sec:results}

Now, let us discuss the numerical implementation of our new treatment of the Schwinger effect during axion inflation. As a benchmark scenario, we choose a model with a simple quadratic potential
\begin{equation}
\label{eq: axion-potential}
    V(\phi) = \frac{m_\phi^2\phi^2}{2}\,
\end{equation}
with mass $m_\phi=\num{6e-6}M_{\mathrm{P}}$. Although the inflationary model with this potential is already ruled out by CMB observations~\cite{Planck:2018jri,BICEP:2021xfz}, it is still a viable effective model that describes inflationary dynamics close to the end of inflation. Another reason for choosing this potential is that it has been used in several previous studies of gauge-field production during axion inflation, both (semi-) analytical~\cite{Domcke:2020zez,Gorbar:2021rlt,Domcke:2023tnn} and on the lattice~\cite{Caravano:2021bfn,Caravano:2022epk,Figueroa:2017qmv,Sharma:2024nfu}. In this sense, this model and the choice of the inflaton mass represent an established benchmark scenario that can be used to compare results across different studies in the literature.

We consider the typical form of the axial coupling function linear in the axion field,
\begin{equation}
    I(\phi) = \frac{\alpha_{\phi}}{f}\,\phi\,,
\end{equation}
where we set $\alpha_{\phi}/f =25/M_{\mathrm{P}}$ in our numerical analysis. Note that we use this value and the quadratic potential in Eq.~\eqref{eq: axion-potential} merely as an example to illustrate the formalism developed in this work. In order to obtain a more comprehensive picture of Schwinger pair production during axion inflation, one would need to perform a parameter scan changing both the potential shape and the axion--vector coupling constant in a reasonable range. This lies beyond the scope of the present paper and will be addressed elsewhere.

To be specific, we associate our Abelian gauge field $A_\mu$ with the SM hypercharge gauge group ${\rm U(1)_{Y}}$. Therefore, in the expressions for the Schwinger conductivities, Eqs.~\eqref{eq: sigma-E-B-gef-mixed} to \eqref{eq: sigma-E-B-gef-magnetic}, we set $e=g'$ and $C = 41/12$, which is equal to half the sum of the cubes of hypercharges of all SM fermions.%
\footnote{The half is needed because Eq.~\eqref{eq: current-colinear-frame} for the Schwinger induced current is written for the case of one Dirac fermion, while the sum is taken over all Weyl fermions of the SM.}
Furthermore, we assume all SM fermions to be massless. This is indeed the case if the Higgs field is stabilized at the origin of field space and the electroweak symmetry is intact. For instance, this can be realized by introducing a non-minimal coupling of the Higgs field to Ricci curvature, $\mathcal{L}\supset \xi_{R}|\Phi|^2 R$, which gives a large mass to the Higgs field during inflation, $m_{\Phi}^2 \propto \xi_{R} H^2$. Moreover, we assume $m_{\Phi}^2\gg g'|E'|$; therefore, the Higgs-field contribution to the Schwinger-induced current is exponentially suppressed and we can only consider the fermionic contributions. Since the energy scale during inflation is typically many orders of magnitude greater than the electroweak scale, one has to take into account the running of the gauge coupling constant to arrive at a realistic description of the Schwinger effect. In our numerical analysis, we use the expression for the coupling constant
\begin{equation}
\label{eq: running-gauge-coupling}
    g'(\mu) = \bigg([g'(m_{Z})]^{-2} + \frac{41}{48\pi^2}\ln\frac{m_Z}{\mu}\bigg)^{-1/2}\, ,
\end{equation}
which takes into account the contributions of all SM particles to the one-loop beta function~\cite{Srednicki:2007book}. Here $m_Z\approx \qty{91.2}{\giga\eV}$ is the $Z$-boson mass, $g'(m_{Z})\approx 0.35$, and the characteristic energy scale for the process of pair creation is taken to be
\begin{equation}
\label{eq: mu-scale}
    \mu = (\rho_E+\rho_B)^{1/4} = \bigg(\frac{\curly{E}{0}+\curly{B}{0}}{2}\bigg)^{1/4}\, .
\end{equation}
For the Schwinger pair-creation scale, $k_{\mathrm{S}}$, we use the expression in Eq.~\eqref{eq: k-S-GEF} with $e=g'(\mu)$.

The initial condition for the inflaton field, $\phi(0)=15.55\,M_{\mathrm{P}}$, and its time derivative, $\dot{\phi}(0) = -\sqrt{2/3}m_\phi M_{\mathrm{P}}$, ensure that the system is on the slow-roll trajectory at least 60 $e$-folds from the end of inflation. The initial values of the gauge-field bilinear quantities $\curly{X}{n}$ and the fermion energy density $\rho_{\chi}$ are all taken to be zero.

The GEF system \eqref{eq: GEF General} represents an infinite chain of equations that has to be truncated to enable a numerical implementation. Following Ref.~\cite{Gorbar:2021rlt}, we use the truncation rule
\begin{equation}
\label{eq: truncation-GEF}
    \curly{X}{n_{\mathrm{cut}}+1} = \bigg(\frac{k_{\mathrm{h}}}{a}\bigg)^2 \curly{X}{n_{\mathrm{cut}}-1}\, , \qquad \mathcal{X}=\mathcal{E},\ \mathcal{B},\ \mathcal{G}\, ,
\end{equation}
where the truncation order $n_{\mathrm{cut}}$ must be taken sufficiently large to ensure that its further increase does not impact the solution, i.e., that the solution of the system is stable against an increase of $n_{\mathrm{cut}}$. The same procedure is applied to truncate the supplementary GEF system for $\tildecurly{X}{n}$, Eq.~\eqref{eq: GEF - Tilde}, with the replacement of $k_{\mathrm{h}}$ by $k_{\mathrm{S}}$ in Eq.~\eqref{eq: truncation-GEF}. The different models considered below require significantly different values of $n_{\mathrm{cut}}$. For models in the electric picture, taking $n_{\mathrm{cut}}=55$ turns out to be enough to achieve a stable GEF solution. The mixed picture requires $n_{\mathrm{cut}}=100$, while the magnetic picture even requires $n_{\mathrm{cut}}=155$.

In addition to solving the GEF system of equations, we also perform a consistency check of the obtained result by employing the mode-by-mode solution in momentum space. For this, we use the GEF solution to compute the coefficients in the mode equation \eqref{eq: Mode Equation, K-dep Sigmas} and then solve it numerically for a discrete set of gauge-field modes. We use these spectra of the produced gauge fields and integrate them over the relevant range of modes to compute the values of the GEF bilinear quantities $\curly{X}{0}$ directly from their definition in Eq.~\eqref{eq: Bilinear Terms Integrals}. The relative difference between the values of $\curly{X}{0}$ found directly from the GEF solution and from the mode-by-mode cross-check described above is used to quantify the consistency of the obtained GEF solution.

\begin{figure}
    \centering
    \includegraphics[width=0.49\textwidth]{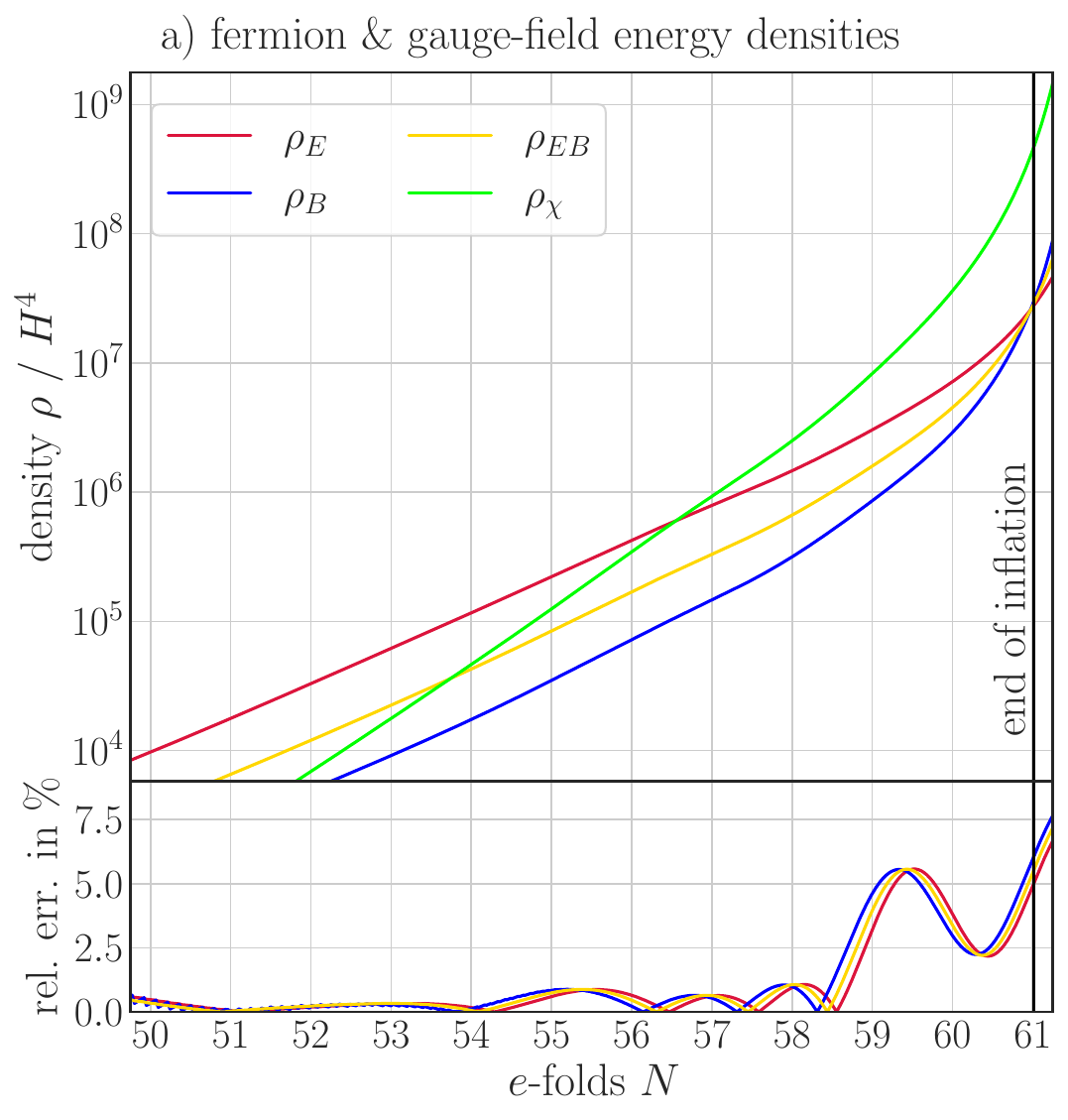}
    \hfill
    \includegraphics[width=0.49\textwidth]{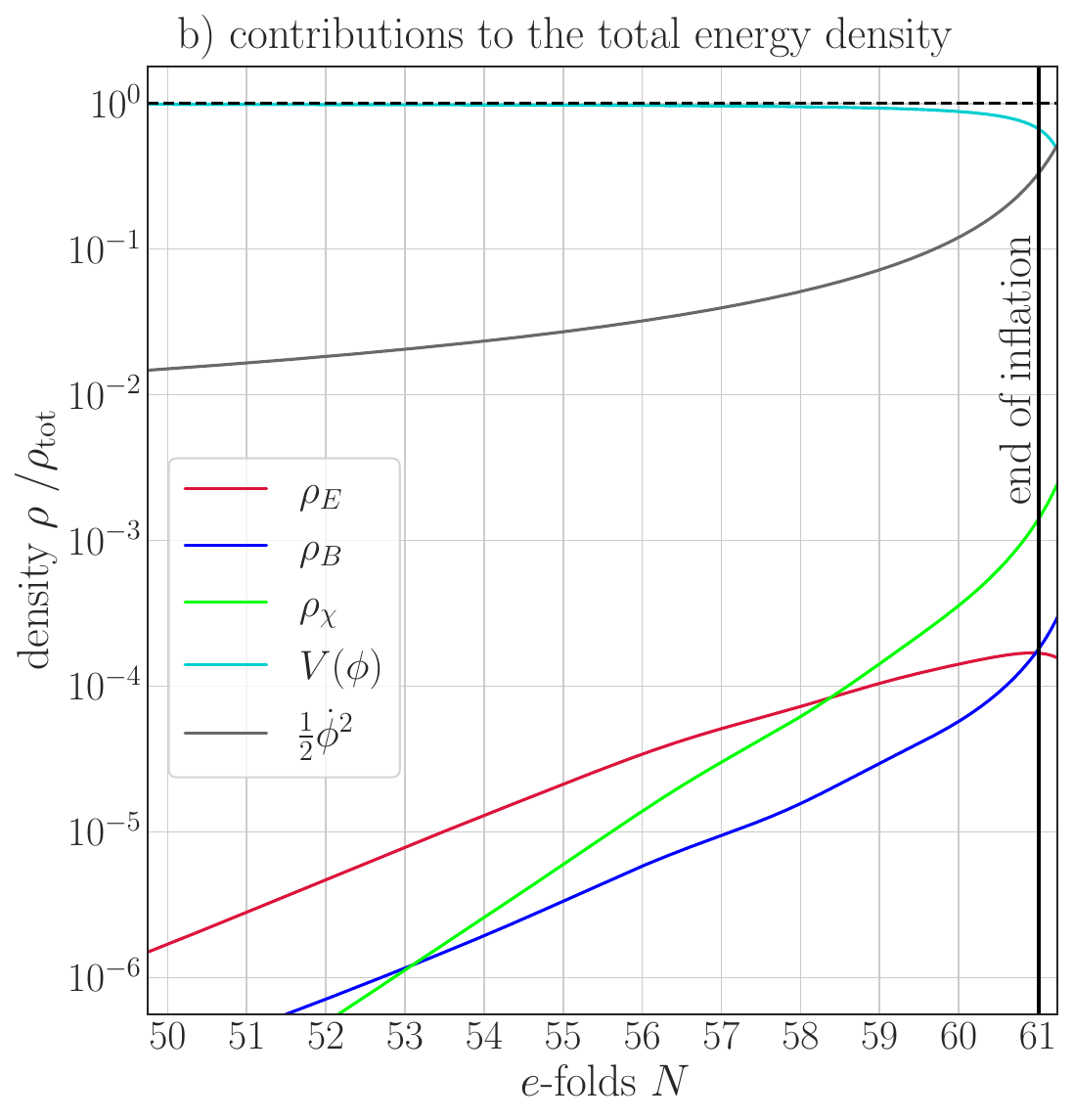}
    \caption{\textbf{Panel~(a)}: The upper part shows the evolution in $e$-folds of the electric energy density $\rho_E$ (red), magnetic energy density $\rho_B$ (blue), the Chern--Pontryagin density $\rho_{EB}$ (yellow), and the fermion energy-density $\rho_{\chi}$ (green), divided by $H^4$. The lower panel shows the relative deviation between the GEF result and the corresponding mode-by-mode solution in momentum space. \textbf{Panel~(b)}: Relative contributions of the electric (red), magnetic (blue) and fermion (green) energy density to the total energy density (dashed black). Also shown are the contributions of the inflaton field to the total energy density stored in kinetic (gray) and potential (cyan) energy. The results in both panels are obtained in the physically motivated model of the Schwinger effect involving the mixed picture for the Schwinger conductivity, Eq.~\eqref{eq: sigma-E-B-gef-mixed}, and the damping on scales $k\leq k_{\mathrm{S}}(t)$, section~\ref{subsubsec: true scale dep}. The dimensionless axion--vector coupling is $\alpha_{\phi} M_{\mathrm{P}}/f=25$.}
    \label{fig:error-mixed}
\end{figure}

The main goal of this paper is to present the improved description of the Schwinger effect during inflation by considering the mixed representation of the induced current in terms of electric and magnetic conductivities, Eq.~\eqref{eq: current-linear-combination}, and by taking into account the scale-dependent Schwinger damping of the gauge-field Fourier modes. Therefore, we start the discussion of our numerical results from the physically most motivated case where the Schwinger conductivities are given by Eq.~\eqref{eq: sigma-E-B-gef-mixed} and the damping is considered for modes with $k\leq k_{\mathrm{S}}(t)$, as described in section~\ref{subsubsec: true scale dep}. The time evolution of the produced gauge fields and fermions during axion inflation in this model is shown in figure~\ref{fig:error-mixed}. Here, the upper part of panel~(a) shows the electric energy density $\rho_E = \langle \bm{E}^2\rangle /2 = \curly{E}{0}/2$ (red), the magnetic energy density $\rho_B = \langle \bm{B}^2\rangle /2 = \curly{B}{0}/2$ (blue), the Chern--Pontryagin density $\rho_{EB}=|\langle \bm{E}\cdot\bm{B}\rangle|/2 = |\curly{G}{0}|/2$ (yellow), and the fermion energy density $\rho_\chi$ (green) in Hubble units. The lower part of panel~(a) presents the relative deviation between the gauge-field densities $\rho_E$, $\rho_B$, and $\rho_{EB}$ obtained from the GEF and the corresponding values computed in the mode-by-mode analysis (the colors of the curves correspond to the quantities shown in the upper panel). Most of the time, the deviation is less than $\qty{1}{\percent}$ and only during the last two $e$-folds of inflation becomes of the order of a few percent. This accuracy (consistency level) of the result should be absolutely sufficient for the majority of physical applications. Panel~(b) shows the relative contributions of $\rho_E$, $\rho_B$ and $\rho_\chi$ to the total energy density, $\rho_{\rm tot} = 3M_{\rm P}^2 H^2$. We also indicate the contributions to $\rho_{\rm tot}$ coming from the inflaton's kinetic energy density, $\frac{1}{2} \dot\phi^2$, in gray, and its potential energy density, $V(\phi)$, in cyan. The total energy density is driven by $V(\phi)$ inducing the near exponential expansion during inflation. Inflation ends when the contributions of the kinetic and potential energy density become of the same order of magnitude. In our benchmark scenario, the fermions and the gauge field contribute negligibly to $\rho_{\rm tot}$ though this may depend on our choice of inflaton potential and coupling strength.

The Schwinger effect appears to be extremely important in this model as it strongly dampens the production of gauge fields. In fact, for the coupling constant $\alpha_{\phi}M_{\mathrm{P}}/f=25$ in the absence of fermions, the gauge field would be strong enough to cause a severe backreaction on the background inflaton evolution, see Refs.~\cite{Domcke:2020zez,Gorbar:2021rlt}. However, in this paper, we do not observe any backreaction at all and the inflaton dynamics remain exactly the same as in the absence of gauge fields and fermions.%
\footnote{We do not claim that Schwinger damping always suppresses the gauge field in such a way that backreaction on the background evolution remains negligible. This is just the case for our benchmark scenario. There may be regions in the parameter space of axion inflation where backreaction occurs even in the presence of Schwinger pair production. The search for this parameter range and the study of the complicated dynamics of the axion, gauge fields, and fermions is a very interesting problem in itself and deserves a separate investigation. We are planning to address this issue in future work.}
The electric energy density dominates over the magnetic one until the very end of inflation where they become approximately equal to each other. This is important in the context of magnetogenesis since, typically, in the models considered in the literature, the magnetic field is subdominant compared to the electric one; see, e.g., Refs.~\cite{Gorbar:2021rlt,Durrer:2023rhc}. Furthermore, the fermion energy density appears to be more than an order of magnitude stronger compared to that of the gauge field at the end of inflation. This may be important for the reheating of the Universe after inflation where the Schwinger effect can be considered as a complementary mechanism~\cite{Tangarife:2017rgl} to the direct inflaton decay and parametric resonance~\cite{Kofman:1997yn,Allahverdi:2010xz,Amin:2014eta}.

\begin{figure}
    \centering
    \includegraphics[width=0.8\textwidth]{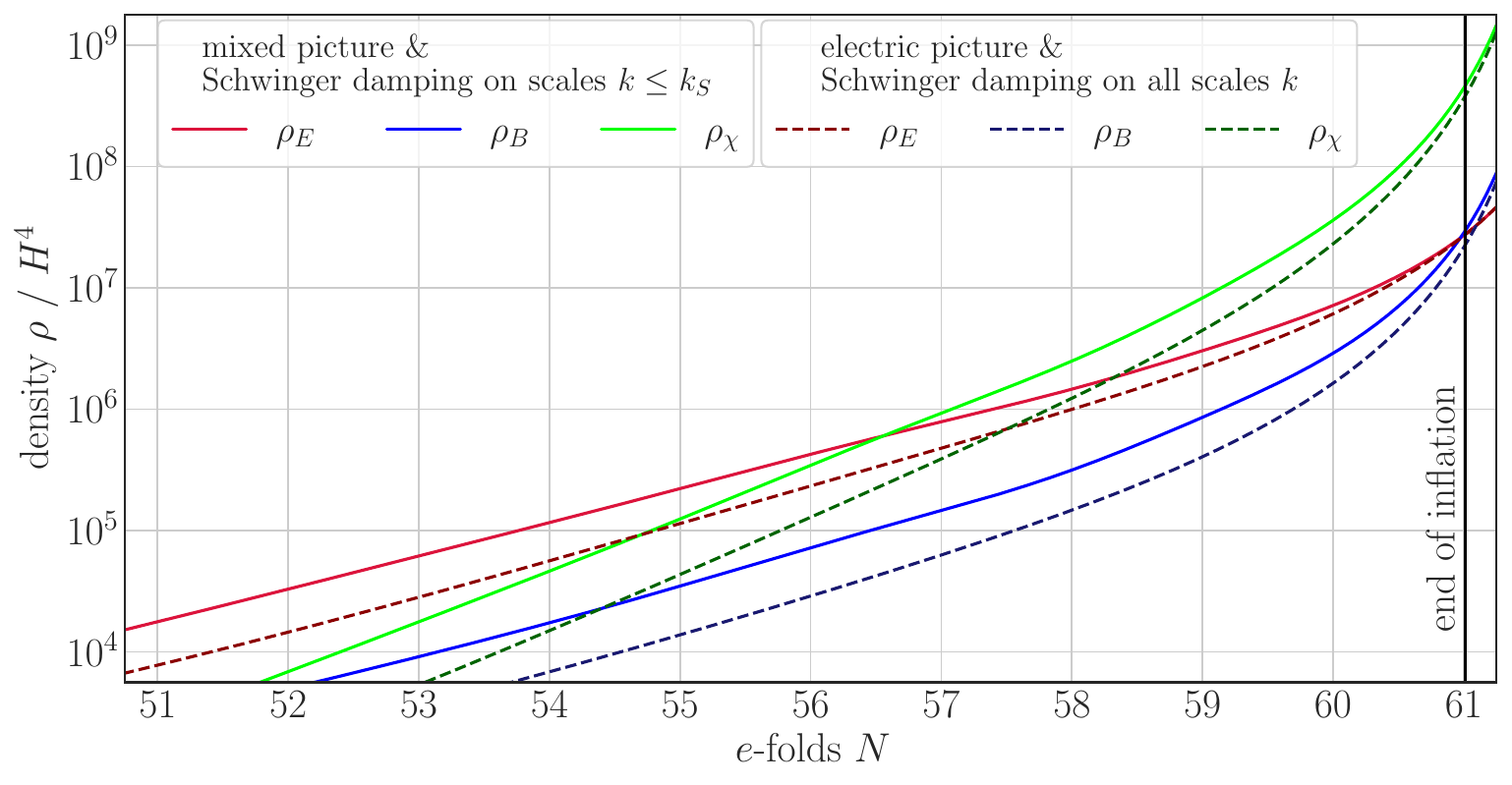}
    \caption{Evolution in $e$-folds of the electric (red), magnetic (blue), and fermion (green) energy densities, divided by $H^4$, obtained in two models of the Schwinger effect: the electric picture with no scale dependence of the Schwinger damping considered in Ref.~\cite{Gorbar:2021rlt} (dark dashed curves) and in the mixed picture with damping on scales $k\leq k_{\mathrm{S}}(t)$ proposed in this work (light solid curves).\label{fig:comparison-old-new}}
\end{figure}

\begin{figure}
    \centering
    \includegraphics[width=0.975\textwidth]{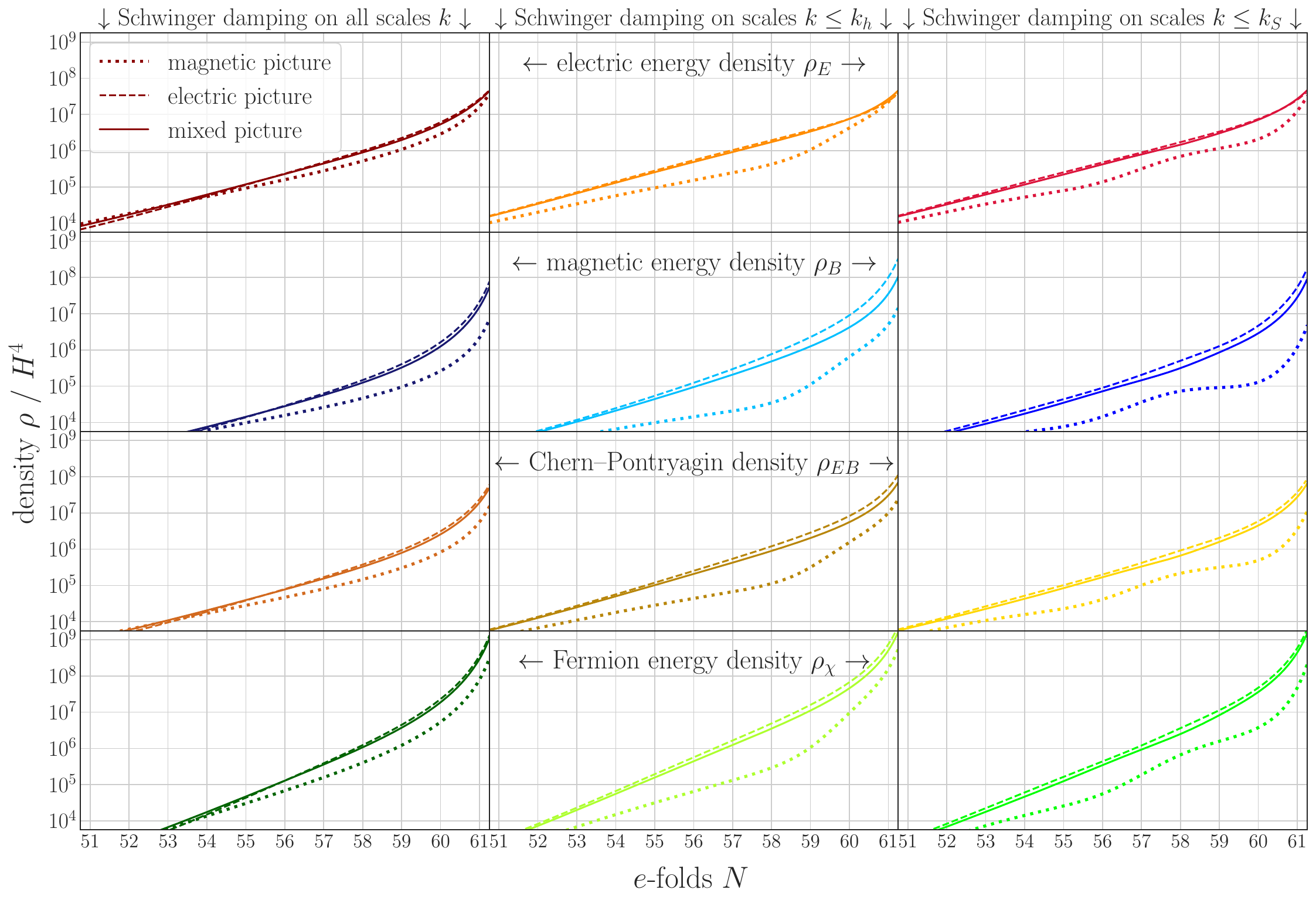}
    \caption{Evolution in $e$-folds of the gauge-field and charged-fermion densities (divided by $H^4$) produced during axion inflation with dimensionless axion--vector coupling $\alpha_{\phi} M_{\mathrm{P}}/f=25$: electric energy density $\rho_E$ (first row, shades of red), magnetic energy density $\rho_B$ (second row, shades of blue), Chern--Pontryagin density $\rho_{EB}$ (third row, shades of yellow), and fermion energy density $\rho_\chi$ (fourth row, shades of green). The first column (dotted lines) corresponds to the magnetic picture for the Schwinger conductivities, Eq.~\eqref{eq: sigma-E-B-gef-magnetic}; the second column (dashed lines) to the electric picture, Eq.~\eqref{eq: sigma-E-B-gef-electric}; and the third column (solid lines) to the mixed picture, Eq.~\eqref{eq: sigma-E-B-gef-mixed}. Each panel compares the results of different models of Schwinger damping: on momentum scales $k\leq k_{\mathrm{S}}(t)$ (section~\ref{subsubsec: true scale dep})\,---\,the lines of the pure color (red, blue, yellow, or green), on all scales (section~\ref{subsubsec: no scale dependence})\,---\,the dark shades of the corresponding color, and on scales $k\leq k_{\mathrm{h}}(t)$ (section~\ref{subsubsec: scale dep kh})\,---\,the light shades of the same color.\label{fig:comparison-damping}}%
\end{figure}

\begin{figure}
    \centering
    \includegraphics[width=0.975\textwidth]{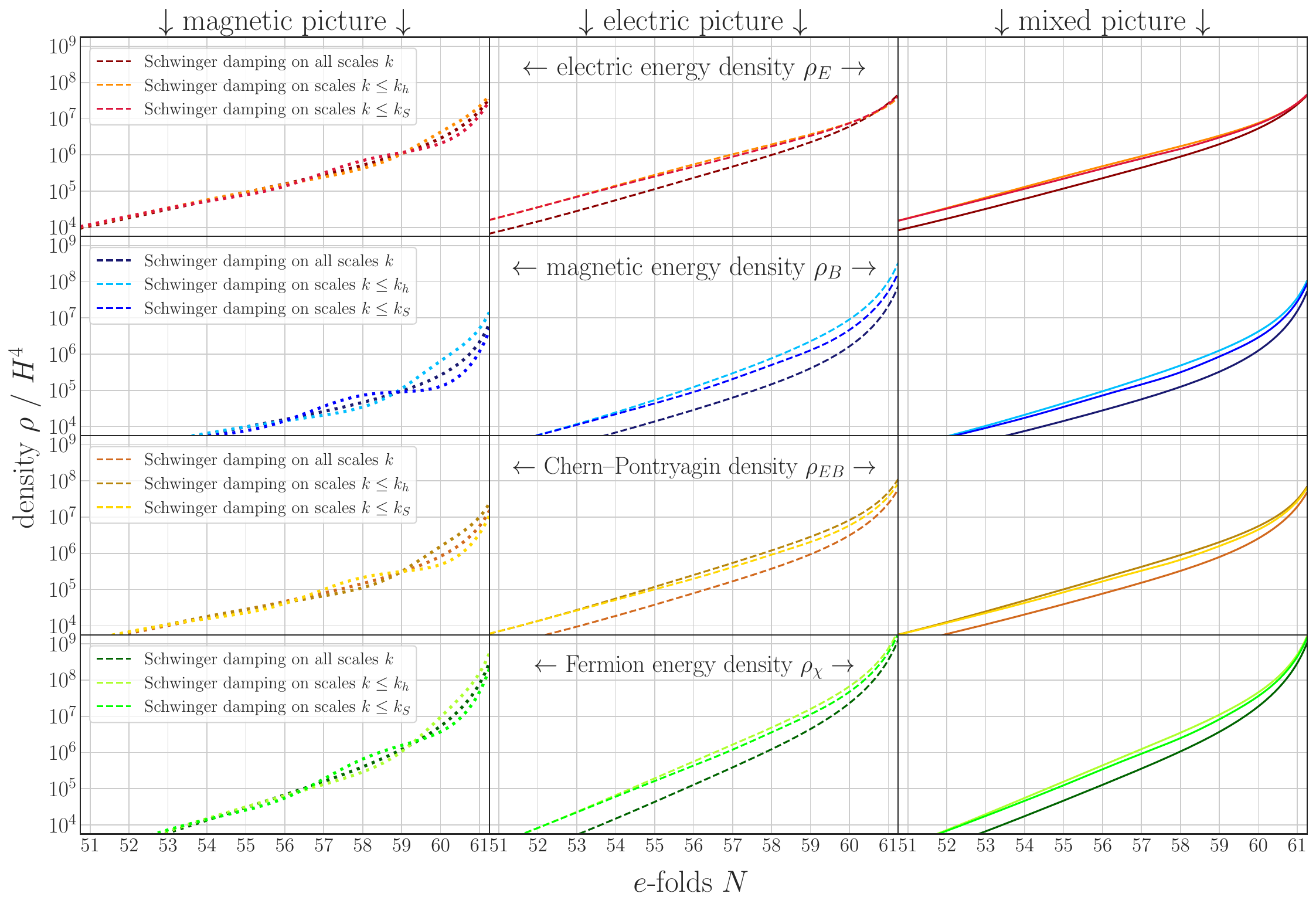}
    \caption{Same results as in figure~\ref{fig:comparison-damping}, rearranged in a different way. The columns now correspond to different models of the Schwinger damping, while each panel compares the results obtained in different pictures for the Schwinger conductivities. \label{fig:comparison-picture}}
\end{figure}

Now, let us compare the results obtained within our new description of the Schwinger effect to the results of already existing approaches in the literature. First, we compare them to the model of Ref.~\cite{Gorbar:2021rlt}, where the electric picture and no scale dependence of the damping was assumed. The corresponding time dependence of the gauge-field and fermion energy densities are shown in figure~\ref{fig:comparison-old-new}. As one can see from this plot, there are significant differences between the predictions of these two models until the very last $e$-folds of inflation, where the corresponding curves (accidentally) converge. Although the final values of the produced gauge-field and fermion densities are roughly the same, the difference between the results far from the end of inflation may still lead to different predictions for the primordial power spectra and, therefore, may be potentially observable. Note that at very early times when the gauge field is too weak to induce Schwinger pair production, both models, of course, give the same results.

Further, let us consider a larger variety of models. In total, we have at hand three representations (pictures) for the induced current and three models of the Schwinger damping described in section~\ref{subsec:different-scale-dependencies}. Combining all of them yields nine different possibilities in total. Numerical results for the produced gauge-field and charged-fermion densities in these nine approaches are shown in Figs.~\ref{fig:comparison-damping} and \ref{fig:comparison-picture}. Here, in both plots, the first row shows the evolution in $e$-folds of the electric energy density $\rho_E$ (red lines), the second row presents the magnetic energy density $\rho_B$ (blue lines), the third row the Chern--Pontryagin density $\rho_{EB}$ (yellow lines), and the fourth row the fermion energy density $\rho_\chi$ (green lines), all divided by $H^4$. Each of the four colors corresponding to the physical quantities listed above is present in three shades, which correspond to three models of Schwinger damping: The pure color (red, blue, yellow, and green) represents the physically motivated model of damping on scales $k\leq k_{\mathrm{S}}(t)$, section~\ref{subsubsec: true scale dep}; the dark shade corresponds to the model with no scale dependence, section~\ref{subsubsec: no scale dependence}; and the light shade is assigned to the case of the simplified model of damping on scales $k\leq k_{\mathrm{h}}(t)$, section~\ref{subsubsec: scale dep kh}. Finally, the line style is associated to the picture for the Schwinger current: the dashed lines correspond to the electric picture, Eq.~\eqref{eq: electric picture}, dotted lines to the magnetic picture, Eq.~\eqref{eq: magnetic picture}, while the solid lines correspond to the mixed one, Eq.~\eqref{eq: mixed picture}. Both figures contain the same curves but arranged in a different way: columns correspond to different pictures while each panel compares the predictions of different models of damping in figure~\ref{fig:comparison-damping} and vice versa in figure~\ref{fig:comparison-picture}.

Comparing the results in different pictures, one observes that the electric and mixed pictures give very close predictions for all observables. This may be explained by the following consideration. During axion inflation, the electric and magnetic fields are in fact very close to the collinear configuration, therefore, $[\curly{G}{0}]^2 \approx \curly{E}{0}\curly{B}{0}$. In this approximation, Eqs.~\eqref{eq: sigma-E-B-gef-mixed} for the Schwinger conductivities in the mixed picture reduce to
\begin{subequations}
\label{eq: sigma-E-B-gef-mixed-collinear}%
    \begin{align}
        \sigma_E &\approx C \frac{\gc^3}{6 \pi^2 H}\frac{|\curly{G}{0}|}{\sqrt{\Sigma}} \sqrt{\frac{\curly{E}{0}}{\curly{E}{0} + \curly{B}{0}}} \coth{\left(\pi \sqrt{\frac{\curly{B}{0}}{\curly{E}{0}}}\right)}\, ,\\
        \sigma_B &\approx -C \frac{\gc^3}{6 \pi^2 H}\frac{\curly{G}{0}}{\sqrt{\Sigma}} \sqrt{\frac{\curly{B}{0}}{\curly{E}{0} + \curly{B}{0}}} \coth{\left(\pi \sqrt{\frac{\curly{B}{0}}{\curly{E}{0}}}\right)}\, .
    \end{align}%
\end{subequations}%
Taking into account also the fact that the electric energy density dominates over the magnetic one until the very end of inflation, the ratio $\curly{E}{0}/[\curly{E}{0}+\curly{B}{0}]\simeq 1$ while $\curly{B}{0}/[\curly{E}{0}+\curly{B}{0}]\ll 1$. Plugging this back into the equations above, we recover exactly the expressions for the conductivities in the electric picture, Eq.~\eqref{eq: sigma-E-B-gef-electric}. Note that, close to the end of inflation, $\curly{E}{0}\approx \curly{B}{0}$ such that $\curly{E}{0}/[\curly{E}{0}+\curly{B}{0}] \simeq \curly{E}{0}/[\curly{E}{0}+\curly{B}{0}] \simeq 1/2$. This is noticeable in our results, where the discrepancy between the electric and mixed picture becomes larger at later times (although still remaining well within one order of magnitude); see, in particular, the second row in figure~\ref{fig:comparison-picture}. The magnetic picture always leads to a larger suppression of the resulting gauge field as compared to the other two pictures. This may be related to the fact that, in the magnetic picture, the conductivity $\sigma_B$ directly modulates the gauge-field production parameter $\xi$ replacing it with $\xi_{\mathrm{eff}}$, which is smaller in absolute value, i.e., it suppresses the tachyonic enhancement of the gauge field. This appears to be more effective than the competition of the tachyonic enhancement by the full $\xi$ and the additional friction from the electric conductivity $\sigma_E$.

Concerning the scale dependence of Schwinger damping, we note that the simplest case where all gauge-field modes are affected by the conductivities irrespectively of their momenta, naturally, leads to the strongest suppression of the produced gauge field and, thus, tends to underestimate the result. The physically motivated model of Schwinger damping on scales $k\leq k_{\mathrm{S}}(t)$ (except for the magnetic picture) predicts larger energy densities for the produced fields by almost one order of magnitude as compared to the case with no scale dependence. Another interesting observation is that the simplified model of damping on scales $k\leq k_{\mathrm{h}}(t)$ leads to results that are extremely close to the physically motivated model of damping; see figure~\ref{fig:comparison-damping}. However, its practical implementation is much simpler and less computationally costly because one does not need to use the additional GEF system \eqref{eq: GEF - Tilde} to compute the quantities $\tildecurly{X}{n}$ and the boundary terms for the main GEF system \eqref{eq: GEF General} do not involve conductivities making them much simpler.


\section{Conclusions}
\label{sec:conclusions}

The production of gauge fields during inflation, in particular, in the axion inflation model, has a lot of phenomenological applications as it may impact the background inflationary dynamics, alter the spectral properties of primordial scalar and tensor perturbations, and give rise to charged particles via the Schwinger effect. The latter phenomenon, in the inflationary context, has been actively discussed in the literature already for a decade and is still rather poorly understood. Even in the simplest setup of de Sitter spacetime with constant and homogeneous electric and magnetic fields (not necessarily collinear to each other), the expression for the electric current of created particles was missing and the mechanism of its backreaction on the gauge field was not appropriately described in the literature. In the present work, we addressed these two issues and implemented an improved description of the Schwinger effect within the framework of the gradient-expansion formalism.

For a generic configuration of the homogeneous electric and magnetic field, we performed a Lorentz boost to the reference frame where the fields become (anti-)collinear to each other. Then, using the known expression for the Schwinger current in this case \cite{Domcke:2018eki,Domcke:2019qmm}, we boosted back to the comoving frame and obtained the vector of the induced current decomposed into the electric- and magnetic-field three-vectors, $\bm{J}=\sigma_E \bm{E} + \sigma_B \bm{B}$, where the conductivities $\sigma_{E/B}$ are non-linear functions of the gauge field given by Eqs.~\eqref{eq: sigma-E-B}. In the case of a quantum gauge field, we then assumed the latter conductivities to be classical functions of the mean gauge field thus keeping the induced current linear in the gauge-field operators. Such a mixed representation for the induced current involving both the electric and magnetic conductivities is the first attempt to (i)~move away from the collinear field approximation and (ii)~remove the existing ambiguity in the description of the Schwinger current which previously was treated in purely electric ($\bm{J}=\sigma_E \bm{E}$) or purely magnetic ($\bm{J}=\sigma_B \bm{B}$) forms.

Furthermore, we qualitatively analyzed the damping of gauge-field modes due to the presence of Schwinger conductivities and identified the characteristic physical momentum scale $k_{\mathrm{S}}$ above which the mode evolution should be insensitive to the presence of the conductive medium. Physically, this is related to the fact that a gauge mode can only be affected by the medium if its wavelength is much larger than the characteristic separation between charge carriers in the plasma. To the best of our knowledge, such a scale-dependent damping of gauge-field modes has not been considered in the literature before.

Further, we incorporated both ingredients discussed above in the GEF. Our new description of the Schwinger effect not only modifies the intrinsic momentum scale of axion inflation\,---\,the tachyonic instability scale $k_{\mathrm{h}}$\,---\,by introducing the conductivities $\sigma_{E/B}$ in its expression, but also introduces a second characteristic scale in the model\,---\,the Schwinger damping scale $k_{\mathrm{S}}$. In practice, this means that we have to double copy the set of the gauge-field bilinear quantities in the GEF. One set handles all gauge-field modes up to $k_{\mathrm{h}}$ and the other one includes a reduced number of modes with momenta $k\leq k_{\mathrm{S}}$. Once $k_{\mathrm{h}}(t) \leq k_{\mathrm{S}}(t)$, this distinction is no longer needed and we are left with only the first type of bilinear quantities. Each set of quantities then requires its own system of equations of motion, which are coupled to each other via the conductivity terms. With this new GEF, we studied the gauge-field and charged-fermion production in a benchmark model of axion inflation with a simple quadratic potential for one particular value of the axion--vector coupling, $\alpha_{\phi}/f=25/M_{\mathrm{P}}$.
Moreover, for comparison, we performed the same computation in other approaches to describe the Schwinger effect already existing in the literature. In total, we considered nine models resulting from the combination of three pictures for the Schwinger-induced current (old electric and magnetic pictures and the new mixed one proposed in this paper) with three models for the Schwinger damping (involving the well-known case of damping on all scales, the newly introduced physically motivated damping on scales $k\leq k_{\mathrm{S}}(t)$ as well as a new approximate model where the damping is assumed to occur only for modes that undergo the tachyonic instability, i.e., on scales $k\leq k_{\mathrm{h}}(t)$).

Our numerical results show that the produced gauge-field and fermion densities are sensitive to the precise modeling of the Schwinger effect. Their values may differ by more than an order of magnitude depending on the chosen approach. Therefore, the improved treatment proposed in this paper, because of its clear physical motivation, must be used in order to obtain accurate numerical results. At the same time, we observed that the predictions obtained within the electric picture are rather close to those in the mixed picture, at least in the benchmark scenario that we studied in this paper. Also, the simplified model where Schwinger damping occurs on scales $k\leq k_{\mathrm{h}}(t)$, being much simpler in realization and computationally less costly (as it does not require doubling the GEF equations), leads to results that are in very good agreement with those for the true model where damping occurs on scales  $k\leq k_{\mathrm{S}}(t)$. Therefore, we recommend using this simplified model for obtaining fast and sufficiently accurate estimates for the energy densities of the produced fields.

The aim of this work was to present a new, improved description of the Schwinger effect during axion inflation. The next step would be to apply this description to the study of gauge-field and fermion production in various realistic models as an intermediate step towards making predictions for observables such as the primordial power spectra, the baryon asymmetry of the Universe, large-scale magnetic fields etc. The imprint of the Schwinger effect in these observables (such as previously considered in Ref.~\cite{Chua:2018dqh} in a different model) is of great interest, as it may provide  evidence for the Schwinger effect operating in the early Universe. We leave this important topic for future studies.

In the end, we would like to note that our new treatment of the Schwinger effect during inflation, although making progress in the understanding of this phenomenon, still does not pretend to be exact. First, it still relies on the expression for the Schwinger-induced current derived in the collinear frame under the assumptions of de Sitter spacetime and constant gauge fields. Second, the local-in-time Ohmic form of the Schwinger current does not take into account the fermion production due to the non-stationary character of the gauge field, the so-called dynamical Schwinger effect, and retardation effects in the fermionic response to the changes in the gauge field. All these issues are assumed to be minor during the slow-roll inflationary phase, even though they may become significant close to the end of inflation. However, the analysis of those effects lies beyond the scope of the present article. We plan to address at least a part of them elsewhere.


\vskip.25cm
\section*{Acknowledgements}
The work of R.\,v.\,E.\ and K.\,S.\ is supported by the Deutsche Forschungsgemeinschaft (DFG) through the Research Training Group, GRK 2149: Strong and Weak Interactions\,---\,from Hadrons to Dark Matter. O.\,S.\ is supported by a Philipp Schwartz fellowship of the University of M\"{u}nster.


\appendix
\section{Properties of the Whittaker functions}
\label{app: Whittaker}

In this appendix, we list the properties of the Whittaker functions that are necessary for the derivation of the boundary terms in the GEF.

The Whittaker differential equation
\begin{equation}
\label{eq: Whittaker-eq}
    \frac{\D^{2}\!f}{\D z^{2}}+\left(-\frac{1}{4}+\frac{\kappa}{z}+\frac{1/4-\mu^{2}}{z^{2}}\right)f=0
\end{equation}
has two linearly-independent solutions, $M_{\kappa,\mu}$ and $W_{\kappa,\mu}$. They are expressed in terms of the Kummer $\Phi$ and Tricomi $U$ confluent hypergeometric functions as follows (see Eqs.~(13.1.32) and (13.1.33) in Ref.~\cite{AbramowitzStegun}):
\begin{align}
    M_{\kappa,\mu}(z)&=e^{-z/2}z^{\mu+1/2}\,\Phi(\mu-\kappa+1/2;1+2\mu;z)\, ,\\
    W_{\kappa,\mu}(z)&=e^{-z/2}z^{\mu+1/2}\,U(\mu-\kappa+1/2;1+2\mu;z)\, .
\end{align}

Using these relations and Eqs.~(13.5.1)--(13.5.2) in Ref.~\cite{AbramowitzStegun}, one can derive the following asymptotic expressions for the Whittaker functions at $|z|\to \infty$ ($-3\pi/2 < \operatorname{arg}z \leq -\pi/2$),
\begin{align}
\label{eq: M-asym}
    M_{\kappa,\mu}(z)&=\frac{\Gamma(2\mu+1)}{\Gamma(\mu+1/2+\kappa)}e^{-i\pi(\mu+1/2-\kappa)}e^{-z/2} z^{\kappa} [1+O(z^{-1})]\nonumber\\
    &+\frac{\Gamma(2\mu+1)}{\Gamma(\mu+1/2-\kappa)}e^{z/2} z^{-\kappa} [1+O(z^{-1})]\, ,\\
\label{eq: W-asym}
    W_{\kappa,\mu}(z)&=e^{-z/2} z^{\kappa} [1+O(z^{-1})]\, .
\end{align}

The derivatives of the Whittaker functions $M$ and $W$ are given by Eqs.~(13.4.32)--(13.4.33) in Ref.~\cite{AbramowitzStegun}:
\begin{align}
\label{eq: Whittaker-M-derivative}
    z\frac{\D}{\D z}M_{\kappa,\mu}(z)&=\Big(\frac{z}{2}-\kappa\Big)M_{\kappa,\mu}(z)+\Big(\frac{1}{2}+\mu+\kappa\Big)M_{\kappa+1,\mu}(z)\, ,\\
\label{eq: Whittaker-W-derivative}
    z\frac{\D}{\D z}W_{\kappa,\mu}(z)&=\Big(\frac{z}{2}-\kappa\Big)W_{\kappa,\mu}(z)-W_{\kappa+1,\mu}(z)\, .
\end{align}
Finally, the Wronskian of the Whittaker functions $W$ and $M$ has the following form:
\begin{equation}
\label{eq: Whittaker-Wronskian}
    \mathcal{W}[W_{\kappa,\mu}(z),\,M_{\kappa,\mu}(z)] = \frac{\Gamma(2\mu+1)}{\Gamma(\mu+1/2-\kappa)}\, ,
\end{equation}
which can be easily derived from the asymptotical expressions in Eqs.~\eqref{eq: M-asym} and \eqref{eq: W-asym} and the fact that the Wronskian is constant.%
\footnote{The Wronskian of two solutions of the linear second-order differential equation $f''+p(z) f' +q(z) f=0$ satisfies equation $\mathcal{W}'+ p(z) \mathcal{W}=0$ which reduced to $\mathcal{W}'=0$ in the case $p(z)=0$ [as, e.g., for the Whittaker equation \eqref{eq: Whittaker-eq}].}

\section{Dynamical evolution of the fermion energy density}
\label{app: rhoChi EoM}

In this appendix, we derive the equation governing the time evolution of the energy density of Schwinger-produced particles. To this end, we first consider the equation of motion for the inflaton energy density $\rho_{\phi}=\dot{\phi}^2/2+V(\phi)$. Using the Klein--Gordon equation \eqref{eq: GEF general - KG}, it is straightforward to obtain
\begin{equation}
    \dot{\rho}_{\phi} + 3H (\rho_{\phi}+p_{\phi}) = -I_{,\phi}\dot{\phi}\,\curly{G}{0}\, ,
    \label{eq: rho-phi-evolution}
\end{equation}
which has the form of a covariant energy-(non)conservation law with $p_{\phi}=\dot{\phi}^2/2-V(\phi)$ being the inflaton pressure and the right-hand side representing the source term that arises from the interaction between the inflaton and the gauge-field components.

Further, we consider the evolution of the gauge-field energy density $\rho_{\mathrm{GF}}=(\curly{E}{0}+\curly{B}{0})/2$. Employing the GEF equations \eqref{eq: GEF General - En} and \eqref{eq: GEF General - Bn} for $n=0$ we get
\begin{equation}
    \dot{\rho}_{\mathrm{GF}} + 3H (\rho_{\mathrm{GF}}+p_{\mathrm{GF}}) = S_{\rho} + I_{,\phi}\dot{\phi}\curly{G}{0} - \big(\sigma_{E}\,\barcurly{E}{0}-\sigma_{B}\,\barcurly{G}{0}\big)\, ,
    \label{eq: rho-GF-evolution}
\end{equation}
where $p_{\mathrm{GF}}=(1/3)\rho_{\mathrm{GF}}$ is the gauge-field pressure and we now have three terms on the right-hand side. The first one,
\begin{equation}
    S_{\rho}\equiv \frac{1}{2}\big(S_{\mathcal{E}}^{(0)}+S_{\mathcal{B}}^{(0)} \big)\, , \label{eq: quantum-source}
\end{equation}
originates from the boundary terms in the GEF equations and describes the rate of change of the gauge-field energy density due to the contributions of new gauge-field modes whose momentum crosses the cutoff scale $k_{\mathrm{h}}$. According to our convention, the corresponding modes evolve from the category of ``quantum fluctuations'' to the category of ``modes with classical behavior'' and start contributing to the bilinear functions $\curly{E}{n}$, $\curly{G}{n}$, and $\curly{B}{n}$. Thus, one may interpret the term \eqref{eq: quantum-source} as the quantum source of the gauge-field energy density. The second term on the right-hand side of Eq.~\eqref{eq: rho-GF-evolution} describes the energy transfer from the inflaton field and has exactly the same expression with the opposite sign compared to the source term in Eq.~\eqref{eq: rho-phi-evolution}. Finally, the third term on the right-hand side of Eq.~\eqref{eq: rho-GF-evolution} describes the gauge-field energy dissipation due to the Schwinger pair production. Therefore, exactly the same term with the opposite sign should appear in the equations of motion for the energy density of the Schwinger-produced particles. Phenomenologically, one can construct this equation as follows:
\begin{equation}
    \dot{\rho}_{\chi} + 3H(\rho_{\chi} + p_{\chi}) = +\big(\sigma_{E}\,\barcurly{E}{0}-\sigma_{B}\,\barcurly{G}{0}\big)\, .
\label{eq: rho-chi-evolution}
\end{equation}
Assuming that the produced particles are ultrarelativistic with a statistically isotropic momentum distribution, we set $p_{\chi}=(1/3)\rho_{\chi}$, which finally brings Eq.~\eqref{eq: rho-chi-evolution} to the form in Eq.~\eqref{eq: GEF General - rho-chi} in the main text.



\providecommand{\href}[2]{#2}\begingroup\raggedright\endgroup


\end{document}